\documentclass[11pt,oneside]{article}
\usepackage[T1]{fontenc}
\usepackage[utf8]{inputenc}

\usepackage{amsmath,amssymb,mathtools}
\usepackage{textgreek}
\usepackage{wrapfig}
\usepackage{tabu}
\usepackage{pstool}
\usepackage{graphicx,epstopdf}
\usepackage{xcolor}
\usepackage{enumerate}
\usepackage{cancel}
\definecolor{lightblue}{rgb}{0.07,0.69,0.84}

\title{Generalized Space-Time Engineered \\ Modulation (GSTEM) Metamaterials}
\date{}

\author{\textbf{C. Caloz$^{*\dagger}$,~\textit{Fellow,~IEEE}, Zoé-Lise Deck-Léger$^\dagger$,}\\ \textbf{A. Bahrami$^*$, O. Céspedes Vicente$^\dagger$ and Zhiyu Li$^{\ddag*}$}\\
\normalsize{\textit{$^*$KU Leuven, Belgium, $^\dagger$Polytechnique Montréal, $^\ddag$Xi'an Jiaotong University}}}

\usepackage[a4paper]{geometry}

\newcommand{\ve}[1]{\mathbf{#1}}

\newcommand{\ves}[1]{\boldsymbol{#1}}
\newcommand{\uve}[1]{\mathbf{\hat{{#1}}}}

\newcommand{\tx}[1]{\text{#1}}
\newcommand{\te}[1]{\overline{\overline{#1}}}

\begin{document}
\maketitle
\begin{abstract}
\small{This article presents a global and generalized perspective of electrodynamic metamaterials formed by space and time engineered modulations, which we name Generalized Space-Time Engineered Modulation (GSTEM) Metamaterials, or GSTEMs. In this perspective, it describes metamaterials from a unified \emph{spacetime viewpoint} and introduces \emph{accelerated metamaterials} as an extra type of dynamic metamaterials. First, it positions GSTEMs in the even broader context of electrodynamic systems that include (non-modulated) moving sources in vacuum and moving bodies, explains the difference between the moving-matter nature of the latter and the moving-perturbation nature of GSTEMs, and enumerates the different types of GSTEMs considered, namely Space EMs (SEMs), Time EMs (TEMs), Uniform Space-Time EMs (USTEMs) and Accelerated Space-Time EMs (ASTEMs). Next, it establishes the physics of the related interfaces, which includes direct-spacetime scattering and inverse-spacetime transition transformations. Then, it exposes the physics of the GSTEM metamaterials formed by stacking these interfaces and homogenizing the resulting crystals; this includes an original explanation of light deflection by USTEMs as being a \emph{spacetime weighted averaging phenomenon} and the demonstration of ASTEM light curving and black-hole light attraction. Finally, it discusses some future prospects. Useful complementary information and animations are provided in the Supplementary Material.}
\end{abstract}

\section{Introduction}\label{sec:intro}

\emph{Metamaterials} are artificial structures consisting of supra-molecular but sub-wavelength particles that are engineered to provide medium properties beyond (\textmu\textepsilon\texttau$\acute{\text{\textalpha}}$) those available in conventional materials~\cite{Caloz_EM_2005,Engheta_MPEE_2006,Capolino_2009}. Following rudimentary ancient nano-composites, medieval stained glasses and XX$^\textrm{th}$ century artificial dielectrics, they have experienced spectacular developments in the past two decades, where they have diversified and expanded to a point that they represent nowadays a powerful new paradigm in science and technology. This evolution has been largely facilitated by the advent of \emph{metasurfaces}~\cite{Bozhevolnyi_MPA_2018,Achouri_EMS_2021}, which may be seen as the two-dimensional counterpart of voluminal metamaterials, with the benefits of easier fabrication, lower loss and greater flexibility, as well as drastic functional extensions of frequency- or polarization-selective surfaces, reflect- or transmit-arrays and spatial light modulators. 

Most of the metamaterials and metasurfaces investigated until recently have been \emph{static}, i.e., modulated only in space; we shall therefore refer to them as Space Engineered Modulation (SEM) metamaterials, or SEMs for short. A major advance in the field has been realized by making metamaterials \emph{dynamic}, either by replacing the space modulation by a time modulation or by adding a time modulation to the space modulation. This introduction of the \emph{dimension of time} as a new structural medium parameter has resulted in the metamaterial classes of Time Engineered Modulation (TEM) metamaterials, or TEMs~\cite{Morgenthaler_1958,Holberg_parametric_1966,Shlivinski_PRL_2018,Mazor_TAP_2020,Pacheco_OPT_2020,Mazor_PRL_2021}, and Space-Time Engineered Modulation (STEM) metamaterials, or STEMs\cite{Cassedy_1963,Biancalana_2007,Yu_APL_2009,Shaltout_OME_11_2015,Hadad_PRB_2015,Chamanara_PRB_10_2017,Deck_APH_10_2019,Caloz_TAP_PI_2019,Caloz_TAP_PII_2019,Huidobro_PNAS_2021}\footnote{The terminology ``time-modulated metamaterials'' and ``space-time modulated metamaterials'' applies to metamaterials that have already a spatial modulation \emph{before} being temporally modulated, but \emph{not} to -- equally relevant! -- metamaterials whose dynamic structure is really \emph{formed} (and \emph{engineered}) by a time or space-time modulation. Hence our introduction of the \emph{general} terms TEMs and STEMs, and related terminology in Tab.~\ref{tab:term_acr}.}. Specifically, TEMs and STEMs are metamaterials that are formed by the variation (modulation) of a medium parameter in time and in both space and time, respectively, induced by an external drive. In the case of electromagnetic metamaterials, on which the paper focuses, the modulated parameter may be the refractive index, the permittivity, the permeability, or any of the bianisotropic and higher-order spatial-dispersion constitutive parameters and combination thereof, while the modulation drive may be acoustic (e.g., surface/bulk acoustic waves in a piezoelectric crystal), electronic (e.g., electric voltage variations in varactor chips), optical (e.g., laser pulses in semiconductor slabs), etc.~\cite{Gaafar_NatP_2019,Shaltout_Science_2019}. TEMs and STEMs may thus be seen as \emph{medium -- generally 3+1D, or 4D -- extensions} of electronic and optical active lumped element and circuit systems, such as parametric amplifiers~\cite{Tien_JAP_1958,Cullen_ProcIEEE_1960} and acousto-electric/optic modulators~\cite{Rhodes_PIEEE_1981,Saleh_Teich_FP_2019}.

This paper presents a global perspective of metamaterials and a generalization of dynamic metamaterials. The \emph{global perspective} consists in describing all metamaterials, including SEMs and TEMs, in terms of \emph{space-time} -- or \emph{spacetime}\footnote{The two spellings (with and without a hyphen) of this word are found in the literature on dynamic systems, whether for the noun or for the adjective. The one-word spelling is the universal standard when referring to the mathematical model that describes the merged nature of the space and time dimensions into a four-dimensional manifold in relativity physics (e.g., curved spacetime), while the spelling with a hyphen is preferable in reference to modulated structures, where the spatial and temporal features of the modulation are distinct and may exist independently of each other (e.g., space-time modulated metasurface). The present paper follows this convention.} -- modulations, with various degrees of complexity, and in connection with the physics of moving bodies, while the \emph{generalization} resides in the extension of STEMs with \emph{uniform} (constant in both space and time) modulation velocity, i.e., Uniform STEMs (USTEM) metamaterials, or USTEMs, to STEMs with \emph{accelerated} modulation, i.e., Accelerated STEM (ASTEM) metamaterials, or ASTEMs. We shall refer to these diverse possible types of metamaterials as Generalized STEMs (GSTEM) metamaterials, or GSTEMs, where it is noted that ASTEMs may feature different orders (derivatives) of acceleration and hence subdivide in further classes. Table~\ref{tab:term_acr} summarizes the terminology.

\begin{table}[h!]
  \centering
  \caption{Terminology and acronyms.} 
  \begin{tabular}{ll}
		\hline\hline
		EM & Engineered Modulation (Metamaterial) \\
		\hline
		SEM & Space EM \\ 
		TEM & Time EM \\ 
		STEM & Space-Time EM \\
		\hline
		USTEM & Uniform(-Velocity) Space-Time EM \\ 
		ASTEM & Accelerated Space-Time EM \\
		\hline
		GSTEM & Generalized Space-Time EM \\
		\hline\hline
  \end{tabular}
  \label{tab:term_acr}
\end{table}
%


\section{Related Electrodynamic Systems}\label{sec:rel_ed_syst}

\begin{figure*}[h!]
	\centering
	\includegraphics[width=\textwidth]{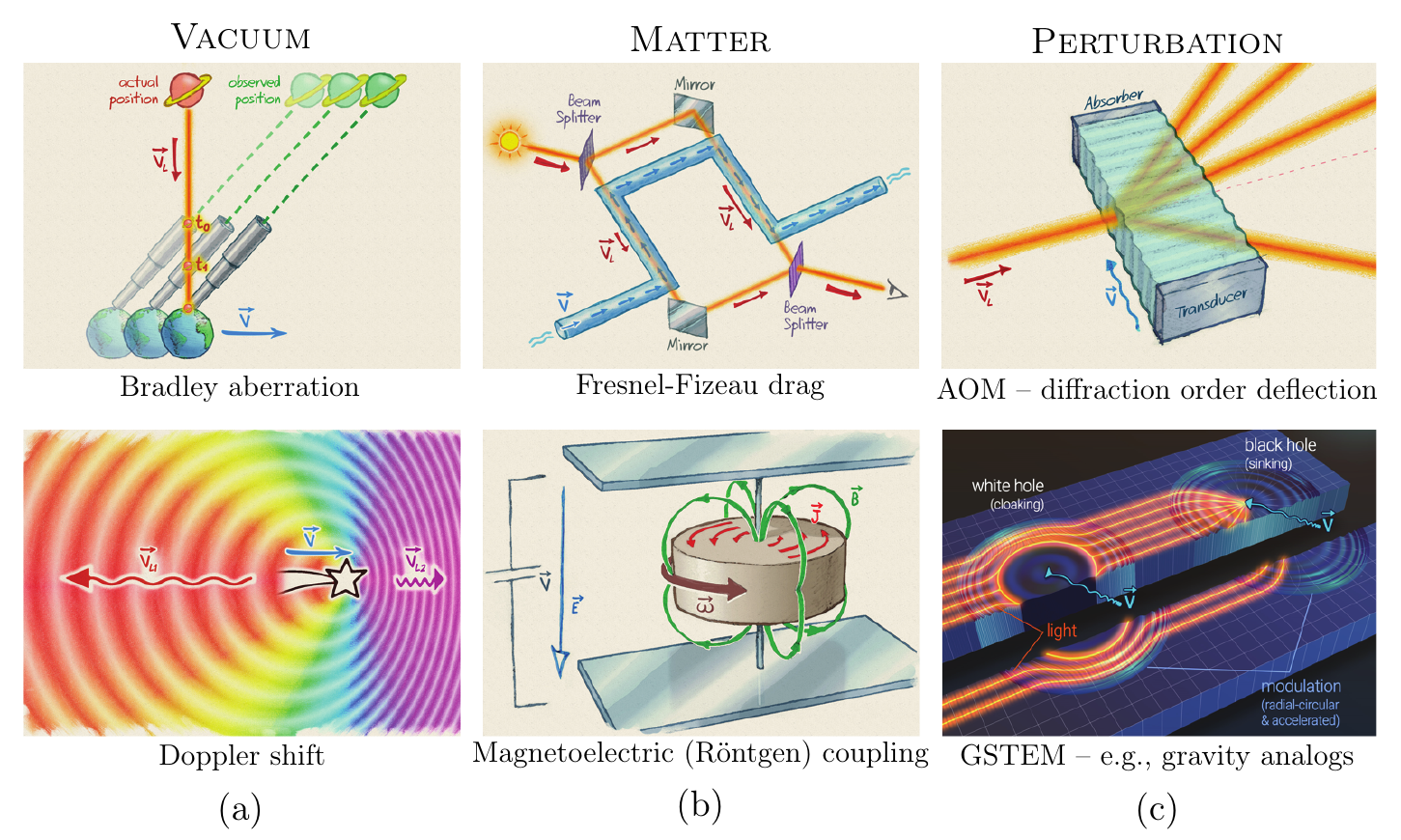}
	\vspace{-6mm}
	\caption{\small Different types of electrodynamic systems and related physical effects. (a)~Moving sources in vacuum. (b)~Moving matter or bodies. (c)~Moving perturbation or modulation.}
	\label{fig:dyn_syst}
\end{figure*}

Electromagnetic GSTEMs are not the only \emph{electrodynamic systems}. They only represent the category of \emph{moving-perturbation} (or moving-modulation) electrodynamic systems. Two other fundamental types of electrodynamic systems should be considered here, \emph{vacuum moving-source} systems and \emph{moving-matter} (or moving-body) systems~\cite{Jackson_1998}, because GSTEMs support physical effects that are inherited from them, although, as we shall see, in distinct embodiments. We shall next describe and compare the three categories, with the help of the illustrations provided in Fig.~\ref{fig:dyn_syst}.

\emph{Vacuum moving-source systems}, illustrated in Fig.~\ref{fig:dyn_syst}(a), are systems involving objects (e.g., star or car) that emit or reflect light\footnote{We use here the term ``light'', as commonly done in the optics community, to designate electromagnetic waves and photons of any frequency or wavelength, for brevity, but a broader spectrum, including radio and terahertz waves, is implicitly assumed.} while moving in \emph{vacuum} relatively to the observer (e.g., Earth or road), with vacuum being defined as a portion of space that is essentially devoid of matter. The earliest reported related effect is \emph{Bradley aberration} (top panel), whereby a terrestrial observer sees a star in a direction that is titled towards the direction of the motion of Earth in its orbit around the sun~\cite{Bradley_1728}. Another effect, which commonly manifests itself in daily life with sound sources, is the \emph{Doppler shift} (bottom panel), whereby an observer of a moving source sees the frequency of the wave emitted or reflected by that source as depending on its velocity, with larger and lower frequency for approaching and receding motion, respectively~\cite{Doppler_1842}. Vacuum-moving-source systems are the simplest electrodynamic systems, since they are restricted to light propagation without light-matter interaction.

\emph{Moving-matter/body systems}, illustrated in Fig.~\ref{fig:dyn_syst}(b), are systems involving \emph{matter} (e.g., water or dielectric) that moves relatively to the observer (e.g., laboratory experimenter) and that supports the propagation of light emitted from the reference frame of the observer, with matter motion defined as a collective translation or/and rotation of atoms and molecules over distances that are much larger than the molecular scale; these systems involve thus typically moving solids, fluids or gases. A related effect is the \emph{Fresnel-Fizeau drag} (top panel), whereby the speed of light is reduced or increased for downstream or upstream propagation in a moving fluid~\cite{Fresnel_1818,Fizeau_1851,Sagnac_1_1913}. Another effect that is of major importance in electrodynamics is the \emph{R\"{o}ntgen magneto-electric coupling} (bottom panel), whereby the motion (here, the rotation) of a solid submitted to an electric field induces a magnetic field in the frame of a rest observer, due to the creation of surface polarization currents~\cite{Rontgen_1888,Sommerfeld_ED_1952}. Moving matter/body systems are more complex than vacuum moving-source systems because of the addition of their matter drag and magneto-electric coupling effects on top of the aberration and Doppler shift effects occurring in vacuum moving-source systems.

Finally, \emph{moving perturbation/modulation systems}, illustrated in Fig.~\ref{fig:dyn_syst}(c), are systems involving a \emph{perturbation} (e.g., an acoustic wave in a piezoelectric crystals) that moves relatively to the observer (e.g., frame of an optical or microwave device) and that scatters light emitted from the reference frame of the observer, with perturbation motion defined as a traveling-wave (or standing-wave) \emph{modulation} of some electromagnetic medium parameter, \emph{without any net transfer of matter}, i.e., with motion restricted to oscillations of bound charges over submolecular distances (dielectric or magnetic polarization)\footnote{Thermodynamics provides an insightful analogy to distinguish moving perturbation and moving matter in associating the former with heat conduction and the latter with heat convection.}. A common example of such a system is the \emph{acousto-optic modulator} (AOM) (top panel), whereby a periodic propagating perturbation (``spacetime modulation grating''), induced by variations of the molecular density of the medium from an electric signal (piezoelectricity), \emph{deflects the diffraction orders} of the incident light in the direction of the perturbation via Bragg-Brillouin scattering~\cite{Rhodes_PIEEE_1981,Saleh_Teich_FP_2019}. \emph{GSTEMs} (bottom panel), particularly USTEMs and ASTEMs, belong to this category of electrodynamic systems, where they generalize acousto-optic-modulator-type systems to multi-dimensional (2+1D = 3D and 3+1D = 4D), multi-velocity (uniform or nonuniform)~\cite{Bahrami_engrXiv_2021}, homogenized~\cite{Huidobro_PNAS_2021} and ``new-physics''\cite{Caloz_TAP_PI_2019,Caloz_TAP_PII_2019} electrodynamic systems.


\section{Moving Perturbation versus Moving Matter}\label{sec:mov_pvm}

\begin{figure*}[h!]
	\centering
	\includegraphics[width=\textwidth]{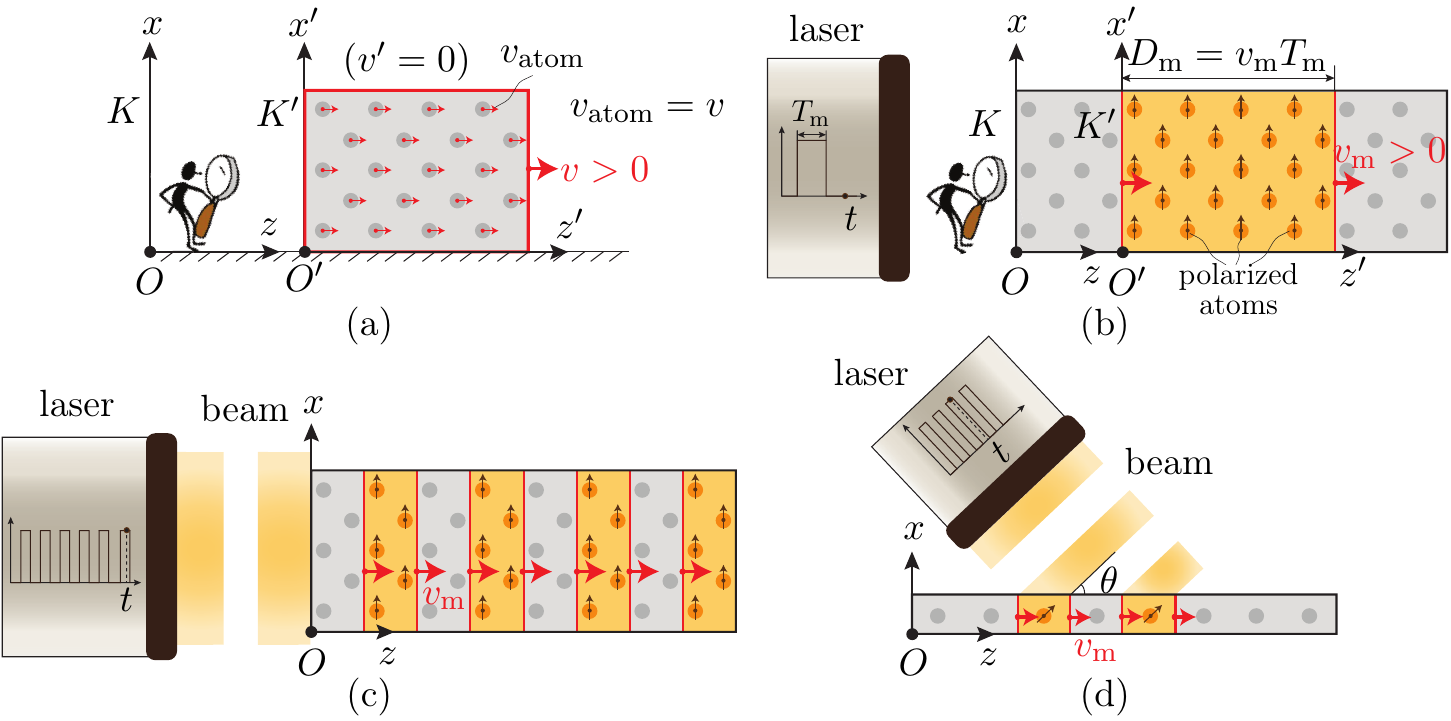}
	\vspace{-6mm}
	\caption{\small Comparison of the electrodynamic structures of the media involved in Figs.~\ref{fig:dyn_syst}(b) and~(c). (a)~Moving (matter) body. (b)~Moving (perturbation) modulation. (c)~Periodic version of (b). (d)~Oblique excitation version of (c), where $v_\mathrm{m}=v_{\mathrm{m}z}=c/\sin\theta$, with even greater diversity attainable with a SEM or STEM mask at the output of the laser. Although the figures consider the specific example of laser-driven modulations, the principle holds for any type of modulation (optical, electronic, acoustic, mechanical, chemical, etc.)~\cite{Gaafar_NatP_2019,Shaltout_Science_2019}.}
\label{fig:comp_mat_pert_str}
\end{figure*}

Figure~\ref{fig:comp_mat_pert_str} compares the electrodynamic structures of the moving-matter/body systems [Fig.~\ref{fig:dyn_syst}(b)] on one hand, and the moving perturbation/modulation systems in [Figs.~\ref{fig:dyn_syst}(c)], which include GSTEMs, on the other hand. Figure~\ref{fig:comp_mat_pert_str}(a) shows a moving-matter system (e.g., sliding curling stone), where the atoms and molecules (matter) move together with the body, along with the comoving frame, $K'$, at a velocity $v$ with respect to the (fixed) laboratory frame, $K$. Figure~\ref{fig:comp_mat_pert_str}(b) shows a moving-perturbation system (e.g., dielectric slab excited by a laser-pump pulse or piezoelectric slab excited by a voltage source). In this case, the atoms and molecules oscillate about their bound position within the (solid) body, under the polarizing effect of the drive excitation, but do not experience any net motion in $K$. Only the related two perturbation interfaces move, inducing a STEM \emph{pulse modulation} of the form $n(z,t)=n_0+\Delta n\cdot\Pi[(z-v_\mathrm{m}t)/D_\mathrm{m}]$, where $n_0$ is the average refractive index, $\Delta n$ is the modulation depth, $\Pi(\cdot)$ is the pulse function, $D_\mathrm{m}$ is the spatial extent of the pulse and $v_\mathrm{m}=v$ is the velocity of the corresponding perturbation. Note that the atoms and molecules, while stationary in $K$, move \emph{with respect to $K'$}, in the \emph{opposite} direction, with velocity $v_\mathrm{atom}'=-v$. 

Figure~\ref{fig:comp_mat_pert_str}(c) shows a continuous (periodic) version of the pulse structure in Fig.~\ref{fig:comp_mat_pert_str}(b) (e.g., using a periodically pulsed laser-pump or electro-acoustic drive), with the STEM medium function $n(z,t)=n_0+\Delta n\cdot\mathrm{sgn}[(\cos(k_\mathrm{m}z-\omega_\mathrm{m}t)+1]/2$, where $\mathrm{sgn}(\cdot)$ is the sign function, and $k_\mathrm{m}$ and $\omega_\mathrm{m}$, with $v_\mathrm{m}=\omega_\mathrm{m}/k_\mathrm{m}$, are the spatial and temporal frequencies, respectively, of the modulation. Finally, Fig.~\ref{fig:comp_mat_pert_str}(d) shows a variant of Fig.~\ref{fig:comp_mat_pert_str}(c), where the drive excites the system from the top of the structure (e.g., oblique laser pump illumination or electro-acoustic source), under an angle $\theta$ with respect to the propagation axis, which provides \emph{superluminal} modulation with velocity $v_\mathrm{m}=v_{\mathrm{m}z}=c/\sin\theta$~\cite{Caloz_TAP_PI_2019}, reducing to instantaneous (pure time) modulation for $\theta=0$.
                                  
Moving perturbation/modulation systems [Figs.~\ref{fig:comp_mat_pert_str}(b)-(d)], and particularly GSTEMs, are more promising than their moving matter/body counterparts [Fig.~\ref{fig:comp_mat_pert_str}(a)] towards real-life applications because i)~they do not require cumbersome moving parts, ii)~they easily attain relativistic velocities and accelerations, and iii)~they posses richer functionality potential, resulting both from their dimensional extension of previous modulated systems and their capability to mimic and transcend cosmological systems (e.g., equivalent horizons and black holes; superluminality and negative mass equivalent)~\cite{Weinberg_GC_1972,MTW_2017}. These are the reasons why GSTEMs are so attractive at this point of research in the field of metamaterials. We shall hereafter restrict our attention to GSTEMs, and refer to the moving matter/body dynamic systems only for the purpose of structural or property comparison.


\section{Perspective and Generalization}\label{sec:presp_gen}

\begin{figure*}[h!]
	\centering
	\includegraphics[width=\textwidth]{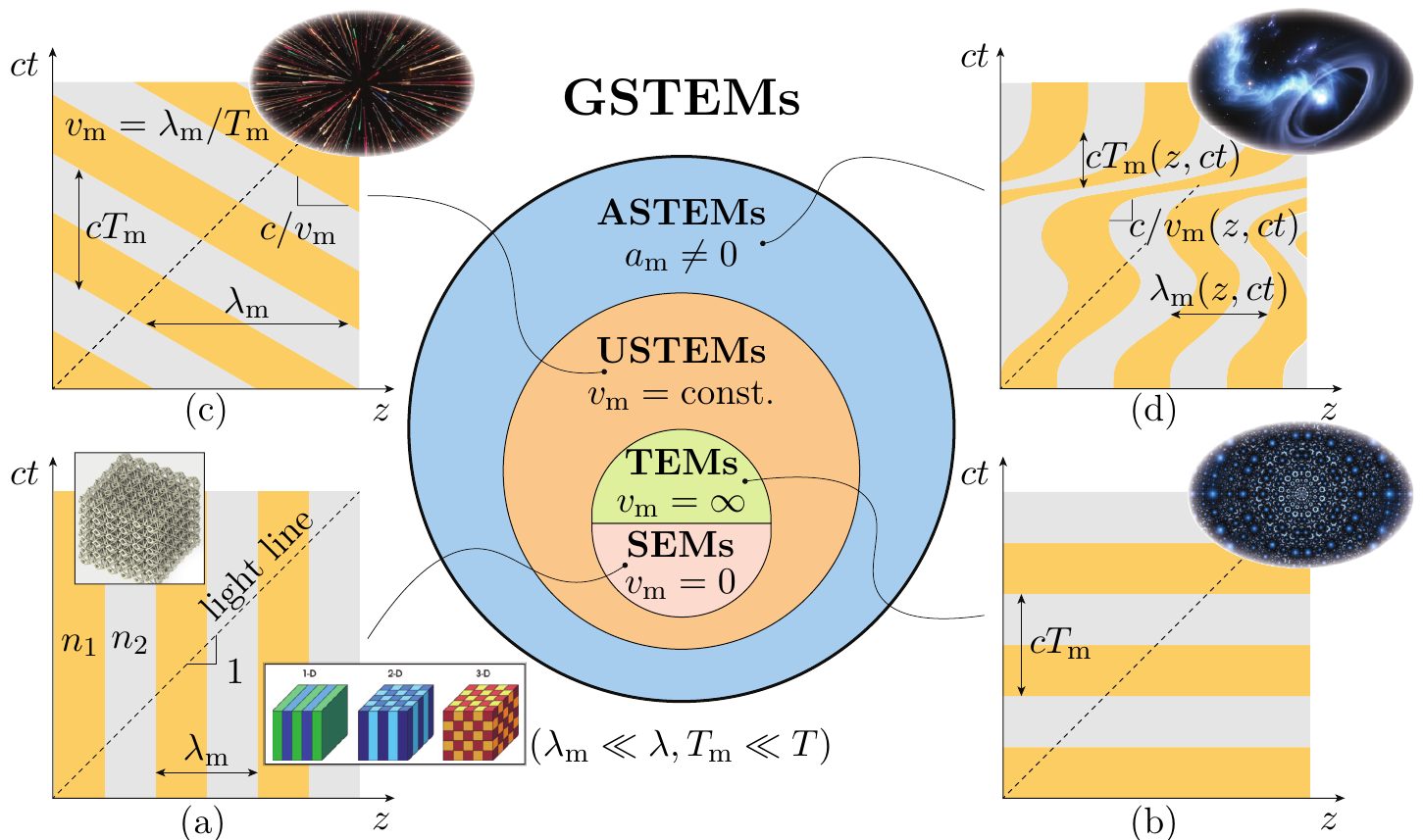}
	\vspace{-4mm}
	\caption{\small Generalized Space-Time (GST) Engineered Modulation (EM) metamaterials -- GSTEMs. (a)~Space EMs -- SEMs. (b)~Time EMs -- TEMs. (c)~Uniform Space-Time EMs (USTEMs). (d)~Accelerated Space-Time EMs (ASTEMs). The subscript `m' refers to `modulation', $n_1$ and $n_2$ are the refractive indices of the constituent media, which are assumed to be isotropic and nondispersive, and $z$ represents the spacetime hyperspace, which may include up to 3 spatial dimensions\ ($x$, $y$, $z$).}
	\label{fig:GSTEMs_persp}
\end{figure*}

Figure~\ref{fig:GSTEMs_persp} depicts the proposed \emph{global perspective} of GSTEMs. The central part of the figure lists the related metamaterials -- SEMs, TEMs, USTEMs and ASTEMs (Tab.~\ref{tab:term_acr}) -- in the order of increasing dynamics generality from the bottom up. The periphery of the figure shows the spacetime (or Minkowski) diagrams~\cite{Minkowski_PZ_1909} corresponding to the four main types of GSTEMs considered in the paper, with suggestive artistic illustrations (Supp. Mat.~\ref{sec:GSTEM_ST_diag}). Such a global perspective offers multiple benefits, including i)~an elegant \emph{classification}, based on the natural concept of spacetime structuration, ii)~a powerful \emph{unification}, suggesting insightful comparisons and cross-fertilization concepts (e.g., time duals of space systems~\cite{Akbarzadeh_OL_2018,Pacheco_OPT_2020,Pacheco_PRB_2021} or space-time extensions of pure space/time systems~\cite{Deck_PRB_2018,Deck_APH_10_2019}), and iii)~a \emph{generalization} of the physics of special relativity~\cite{Einstein_1905,Pauli_TR_1958,French_SR_1968} and general relativity~\cite{Einstein_1915,Pauli_TR_1958,Carroll_SG_2019} (Supp. Mat.~\ref{sec:rel_princ}), which rather involves sources in vacuum [Fig.~\ref{fig:dyn_syst}(a)] or moving bodies [Fig.~\ref{fig:dyn_syst}(b)].

The concept of a \emph{continuous medium}\cite{Landau_ECM_2013} is an idealization. In reality, all materials are formed by a more or less (crystal or amorphous) periodic collection of particles -- atoms and molecules in the case of conventional materials and resonant scatterers in the case of metamaterials -- which subtend the macroscopic response of the medium in terms of dipolar/multipolar responses at the microscopic scale. All materials can therefore be represented as periodically alternating regions of vacuum and particles, as suggested by the alternating gray-golden bands in the spacetime diagrams of Fig.~\ref{fig:GSTEMs_persp}. The case of GSTEMs seems a priori more complicated, because the modulation can take diverse and complex forms. Consider for instance, the sinusoidal modulation, $n(z,t)=n_0+\Delta n\cos(k_\mathrm{m}z-\omega_\mathrm{m}t)$, which is common in practice. Even though this modulation continuously varies, the wave propagating in the medium is myopic to such fine detail in the \emph{metamaterial -- sub-wavelength and sub-period -- regime}, corresponding to the twofold condition $\lambda_\mathrm{m}\ll\lambda$ and $T_\mathrm{m}\ll T$ ($\lambda$ and $T$: smallest wavelength and period of the wave; $\lambda_\mathrm{m}=2\pi/k_\mathrm{m}$ and $T_\mathrm{m}=2\pi/\omega_\mathrm{m}$); it only probes index extrema, with blurred transitions between them, and the modulation can therefore be safely approximated by the discrete binary function $n(z,t)=n_0+\Delta n\cdot\mathrm{sgn}[(\cos(k_\mathrm{m}z-\omega_\mathrm{m}t)+1]/2$ [Sec.~\ref{sec:mov_pvm} and Fig.~\ref{fig:comp_mat_pert_str}(c)]\footnote{Although the paper focuses on the metamaterial (i.e., homogeneous) regime, the concepts of GSTEMs naturally extend to the Bragg regime, where the GSTEM structures, better called then GSTEM \emph{crystals}, exhibit interesting oblique band-gap configurations and physical properties~\cite{Cassedy_1963,Chamanara_PRB_10_2017,Deck_APH_10_2019}.}. Thus, the spacetime diagrams of GSTEMs can generally be represented by the periodic bilayer spacetime patterns shown in Fig.~\ref{fig:GSTEMs_persp}.

SEMs [Fig.~\ref{fig:GSTEMs_persp}(a)] are GSTEMs whose parameters vary only in space. They represent the particular \emph{static} limit case of GSTEMS with zero modulation velocity, $v_\mathrm{m}=0$, and include conventional metamaterials~\cite{Caloz_EM_2005,Engheta_MPEE_2006,Capolino_2009} as well as photonic crystals in the long-wavelength regime~\cite{Joannopoulos_PC_2008}. TEMS [Fig.~\ref{fig:GSTEMs_persp}(b)] are GSTEMs whose features do not vary in space but in time, as some parametric amplifiers~\cite{Tien_JAP_1958} and solid-state time crystals~\cite{Wilczek_PRL_2012_C,Wilczek_PRL_2012_Q}. They represent the particular \emph{instantaneous} limit case of GSTEMS with infinite modulation velocity, $v_\mathrm{m}=\infty$~\cite{Morgenthaler_1958,Holberg_parametric_1966,Shlivinski_PRL_2018,Mazor_TAP_2020,Mazor_PRL_2021}. We have next actual STEMs, with modulation occurring both in space and in time. USTEMs [Fig.~\ref{fig:GSTEMs_persp}(c)] are characterized by a constant modulation velocity, $v_\mathrm{m}=\mathrm{const.}$, as typical acousto-optical modulators (top panel of Fig.~\ref{fig:dyn_syst}(c))~\cite{Cassedy_1963,Biancalana_2007,Yu_APL_2009,Shaltout_OME_11_2015,Hadad_PRB_2015,Chamanara_PRB_10_2017,Shaltout_Science_2019,Deck_APH_10_2019,Caloz_TAP_PI_2019,Caloz_TAP_PII_2019,Huidobro_PNAS_2021}, while ASTEMs [Fig.~\ref{fig:GSTEMs_persp}(d)], introduced only recently~\cite{Bahrami_engrXiv_2021}, are characterized by an accelerated modulation, $a_\mathrm{m}\neq 0$, which may be constant in the moving frame (constant proper acceleration), or have nonzero temporal derivatives (jerk, snap, crackle, pop, etc.)~\cite{Harris_SVH_2002}.


\section{Interface Physics}\label{sec:int_phys}

As illustrated by the spacetime diagrams in Fig.~\ref{fig:GSTEMs_persp}, GSTEMs can be modeled by alternating isotropic medium layers. The \emph{interfaces} delimiting these layers are therefore the main discontinuities or non-uniformities of the structure and represent hence the entities that underpin the light-matter interaction of the metamaterial. For this reason, this section focuses on GSTEM interfaces, while Sec.~\ref{sec:mtm_phys} will reveal how the related principles extend to complete GSTEM media. 

\begin{figure}[h!]
    \begin{minipage}{0.495\textwidth}
        \includegraphics[width=1\textwidth]{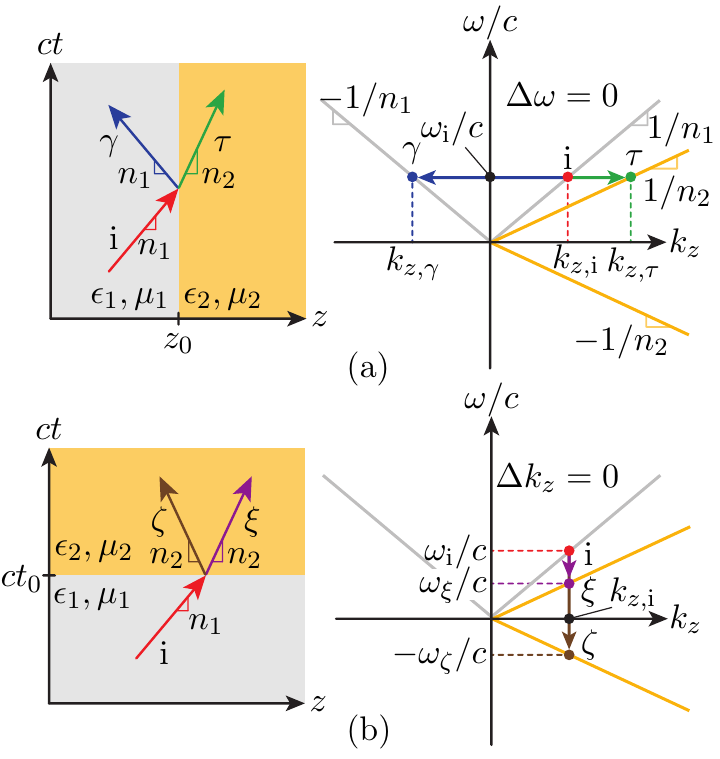}
	    \caption{\small Electrodynamics of the simplest GSTEM interfaces, represented in terms of direct (left) and inverse (right) spacetime diagrams, for the case of normal incidence. (a)~SEM interface [Fig.~\ref{fig:GSTEMs_persp}(a)]. (b)~TEM interface [Fig.~\ref{fig:GSTEMs_persp}(b)].}
	   \label{fig:SEM_TEM_interf}
    \end{minipage}
    \hspace{0.01\textwidth}
    \begin{minipage}{0.495\textwidth}
        \includegraphics[width=1\textwidth]{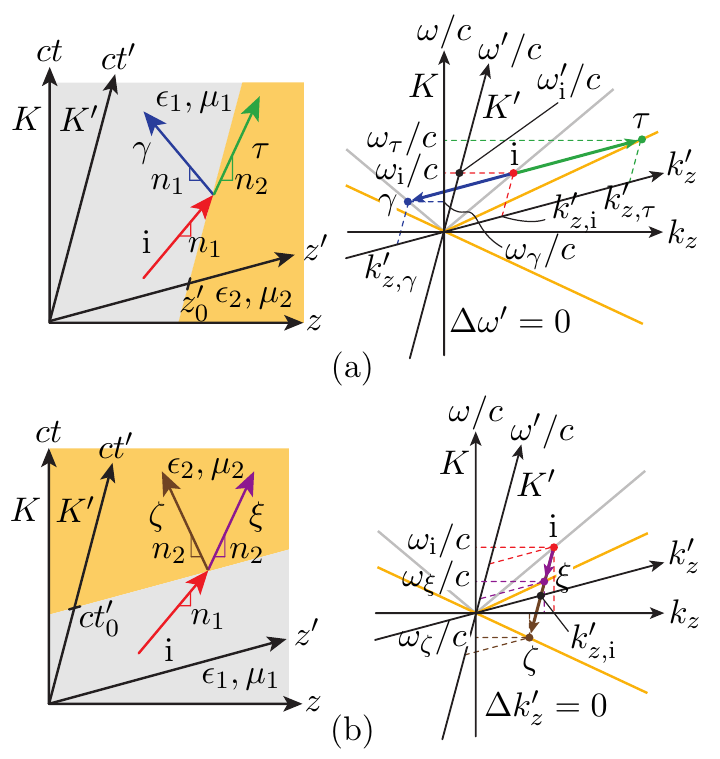}
	    \vspace{-2mm}
	    \caption{\small Electrodynamics of USTEM interfaces [Fig.~\ref{fig:GSTEMs_persp}(c)], represented in terms of direct (left) and inverse (right) spacetime diagrams, for the case of normal incidence. (a)~Subluminal (space-like) regime. (b)~Superluminal (time-like) regime.}
	    \label{fig:STEM_interf}
    \end{minipage}
\end{figure}

Let us start with the simplest cases of SEM and TEM interfaces. The electrodynamics of these interfaces is described in Fig.~\ref{fig:SEM_TEM_interf}, with Fig.~\ref{fig:SEM_TEM_interf}(a) and (b) representing the SEM and TEM cases, respectively~\cite{Caloz_TAP_PII_2019}. When a wave hits a simple interface, or \emph{SEM interface}, it splits into a reflected wave and a transmitted wave, which propagate in opposite directions over time, with well-known scattering (Fresnel) coefficients $\gamma$ and $\tau$, as shown in the left panel of Fig.~\ref{fig:SEM_TEM_interf}(a). These coefficients are found by enforcing the conservation of the tangential $\ve{E}$ and $\ve{H}$ fields at the (spatial) interface discontinuity, which is required to avoid making the $\ve{B}$ and $\ve{D}$ fields singular (at the interface) through the spatial derivative ($\ves{\nabla}\times$) in Maxwell equations~\cite{Caloz_TAP_PII_2019}. On the other hand, the transmitted wavelength is compressed (or the wavenumber increases) if the second medium is denser, as shown in the right panel of  Fig.~\ref{fig:SEM_TEM_interf}(a). Note that this transformation does not involve any change of temporal frequency ($\Delta\omega=0$, energy conservation) since the discontinuity is purely spatial.

The problem of a simple instantaneous interface, or \emph{TEM interface}, is the perfect dual of that of the SEM interface. Now the incident wave splits into a later backward wave and a later forward wave, which also propagate in opposite directions, but in the same (later) medium and with different scattering coefficients \cite{Morgenthaler_1958}, $\zeta$ and $\xi$, as shown in the left panel of Fig.~\ref{fig:SEM_TEM_interf}(b). These coefficients are found by enforcing the conservation of $\ve{B}$ and $\ve{D}$ at the (temporal) interface discontinuity, which is required to avoid making $\ve{E}$ and $\ve{H}$ singular (at the interface) through the temporal derivative ($\partial/\partial t$) in Maxwell equations~\cite{Caloz_TAP_PII_2019}. The two scattered waves are red-shifted (or their frequency decreases) if the second medium is denser, as shown in the right panel of  Fig.~\ref{fig:SEM_TEM_interf}(b). It is thus the temporal frequency that changes in this transformation, while the spatial frequency remains unchanged ($\Delta k_z=0$, momentum conservation), since the discontinuity is purely temporal. This space-to-time duality has several applications, including the \emph{inverse prism}, a device that, instead of decomposing colors into angles as the Newton prism, maps angles into colors~\cite{Akbarzadeh_OL_2018}.

STEM interfaces are even more interesting than their SEM and TEM counterparts. The electrodynamics of USTEM interfaces is described in Fig.~\ref{fig:STEM_interf}, with Figs.~\ref{fig:STEM_interf}(a) and (b) representing the \emph{subluminal} regime [$v_{\mathrm{m}z}<c/\max(n_1,n_2)$, e.g., as in Fig.~\ref{fig:comp_mat_pert_str}(c)] and the \emph{superluminal} regime [$v_{\mathrm{m}z}>c/\min(n_1,n_2)$, e.g., as in Fig.~\ref{fig:comp_mat_pert_str}(d)], respectively~\cite{Caloz_TAP_PII_2019}. Now the modulation occurs simultaneously in space and time, as in all the electrodynamics systems represented in Fig.~\ref{fig:dyn_syst}. Scattering is space-like -- with reflected and transmitted waves -- for the subluminal case, and time-like -- with later backward and later forward waves -- in the superluminal case~\cite{Caloz_TAP_PII_2019}. The corresponding scattering coefficients are found by enforcing the continuity of $(\ve{E}',\ve{H}')$ in the subluminal comoving frame ($K'$, where $\Delta\omega'=0$) and of $(\ve{D}',\ve{B}')$ in the superluminal simultaneity frame ($K'$, where $\Delta k_z'=0$)~\cite{Deck_PRB_2018}, which reveals, upon inverse-Lorentz transformation~\cite{Lorentz_1904,French_SR_1968,Vanbladel_RE_1984} (Supp. Mat.~\ref{sec:Lorentz_transf}), the conservation of the quantities $E_x-v_\mathrm{m}B_y$ and $B_y-v_\mathrm{m}E_x/c^2$ for the plane wave with components $(E_x,B_y,k_z)$ and of $E_y+v_\mathrm{m}B_x$ and $B_x+v_\mathrm{m}E_y/c^2$ for $(E_y,-B_x,k_z)$~\cite{Kong_1972,Vanbladel_RE_1984,Kong_EWT_2008,Caloz_TAP_PII_2019}. On the other hand, the spectral transitions, whose reflective and transmissive parts are manifestations of the \emph{Doppler effect} and of an \emph{index contrast effect}, respectively, are oblique, since the discontinuity is both spatial and temporal, leading to simultaneous spatial and temporal frequency transformations~\cite{Chamanara_TAP_2019}. In the case of \emph{oblique incidence}, the dispersion curves, between which the oblique transitions occur, are altered by the change of the incident momentum on the interface and the scattered angles are deflected towards/against the direction of motion in the sub/super-luminal cases~\cite{Deck_PRB_2018} (Supp. Mat.~\ref{sec:obl_inc}).

Finally, an ASTEM interface may be seen as a generalization of a USTEM interface, where both the direct-spacetime interfaces and the normal-incidence dispersion lines change from straight to curved, as illustrated in Fig.~\ref{fig:GSTEMs_persp}(d) for the case of an ASTEM metamaterial with complex acceleration profile, including direction reversal, and hence jerk ($\partial a_\mathrm{m}/\partial t\neq{0}$). While an accelerated system is \emph{locally} uniform, so that special relativity and Lorentz transformation apply \emph{locally}, it globally requires a much more complex treatment that belongs to the realm of general relativity and, hence, differential geometry~\cite{Sher_HGT_2001,Tu_IM_2011,Carroll_SG_2019}. In this case, the (linear) Lorentz transformations must be replaced by nonlinear transformations corresponding to the type of acceleration at hand~\cite{MTW_2017,Carroll_SG_2019}, the simplest being the $K'$-constant and rectilinear (or proper) acceleration, which is associated with Rindler transformations (Supp. Mat.~\ref{sec:Rindler_transf}). Physically, the USTEM space-time transformation (diffraction-Doppler) effect is promoted to a \emph{space-time chirping effect}~\cite{Tanaka_PRA_1982,Leonhardt_PRA_1999}, with still little explored physics and application potential. A first application is the recently reported ASTEM waveform generator, whose properly design acceleration trajectory allows virtually arbitrary pulse shaping~\cite{Bahrami_engrXiv_2021}.


\section{Metamaterial Physics}\label{sec:mtm_phys}

Stacking the different GSTEM interfaces described in Sec.~\ref{sec:int_phys} leads to the formation of the corresponding GSTEM structures in Fig.~\ref{fig:GSTEMs_persp} [Fig.~\ref{fig:SEM_TEM_interf}(a) $\rightarrow$ Fig.~\ref{fig:GSTEMs_persp}(a), Fig.~\ref{fig:SEM_TEM_interf}(b) $\rightarrow$ Fig.~\ref{fig:GSTEMs_persp}(b), Fig.~\ref{fig:STEM_interf} $\rightarrow$ Fig.~\ref{fig:GSTEMs_persp}(c), curved version of Fig.~\ref{fig:STEM_interf} $\rightarrow$ Fig.~\ref{fig:GSTEMs_persp}(d)]. Before homogenization, these structures are generally periodic spacetime structures (or, possibly, only locally quasi-periodic with a period gradient for the case of ASTEMs), or \emph{GSTEM crystals}, and they involve the same direct (scattering) and inverse (transition) spacetime transformations as their interface counterparts. However, these crystals represent \emph{spacetime-extended structures}, as opposed to spacetime-localized discontinuities, and support in addition multiple spacetime scattering and multiple spacetime transitions, which leads to specific spacetime crystal properties~\cite{Deck_APH_10_2019}.

The GSTEM crystal can be \emph{spatially} 1D, 2D or 3D, as shown in the bottom right inset of Fig.~\ref{fig:GSTEMs_persp}(a). If the wave of interest propagates in a space of dimension that is larger than the spatial dimension of the crystal [specifically, oblique incidence in a 1D crystal and off-plane incidence in a 2D crystal], the crystal is seen by the wave as anisotropic, with different tangential field components for different (e.g., p and s) wave polarizations. Since this represents the most general and most common configuration in early GSTEMs, let us consider henceforth this case of an \emph{anisotropic crystal}, noting in passing that anisotropy, with an anisotropic medium being defined as a medium that exhibits different properties in different directions, is a purely spatial concept, since time is monodimentional. 

In moving-matter dynamic systems [e.g., Fig.~\ref{fig:comp_mat_pert_str}(a)], the problems are ideally solved in the comoving frame ($K'$), where everything is stationary, and hence the boundary conditions and the constitutive relations are both trivial. The situation is more complicated in moving-perturbation systems [e.g., Fig.~\ref{fig:comp_mat_pert_str}(b)], because they involve motion in \emph{both} frames [e.g., moving interfaces in $K$ and (backward) moving atoms and molecules in $K'$]! Let us still choose the $K'$ frame, on the ground that moving matter with stationary boundaries is a known problem~\cite{Deck_Photon_04_2021}, whereas moving boundaries is a more difficult problem. To enter the metamaterial regime, we homogenize the GSTEM by prescribing subwavelength [$\max(\lambda_\mathrm{m})\ll\min(\lambda)$] and subperiod [$\max(T_\mathrm{m})\ll\min(T)$] operation. This operation leads to GSTEM-metamaterial anisotropic constitutive parameters with the same tensorial structure as that of their crystal parent~\cite{Huidobro_PNAS_2021,Bahrami_engrXiv_2022}. However, another effect is present in $K'$: the contradirectional motion of polarized atoms and molecules [e.g., focus on the moving $K'$ frame with $K$-stationary atoms and molecules in Fig.~\ref{fig:comp_mat_pert_str}(b)] ($v_\mathrm{atom}'=-v$); the problem is thus a \emph{moving-matter} problem in $K'$, with the usual \emph{Fresnel-Fizeau drag} and related \emph{motion bianisotropy} [i.e., R\"{o}ntgen magnetoelectric coupling, bottom panel of Fig.~\ref{fig:dyn_syst}(b) (Supp. Mat.~\ref{sec:mov_mat_const_rel} and~\ref{sec:Mov_Mat_disp_rel}), due to the magnetic part of the Lorentz force, $\ve{F}=q\ve{E}+q\ve{v}\times\ve{B}$]~\cite{Kong_1972,Vanbladel_RE_1984} that is associated with the magnetic part of the Lorentz force (Supp. Mat.~\ref{sec:force_bian_inf}). GSTEM metamaterials are thus generally \emph{bianisotropic} in $K'$~\cite{Kong_1972,Kong_EWT_2008,Caloz_APM1_02_2020,Caloz_APM2_04_2020} (Supp. Mat.~\ref{sec:Mov_Mat_disp_rel}).

\begin{figure}[h!]
    \begin{minipage}{0.669\textwidth}
        \includegraphics[width=1\textwidth]{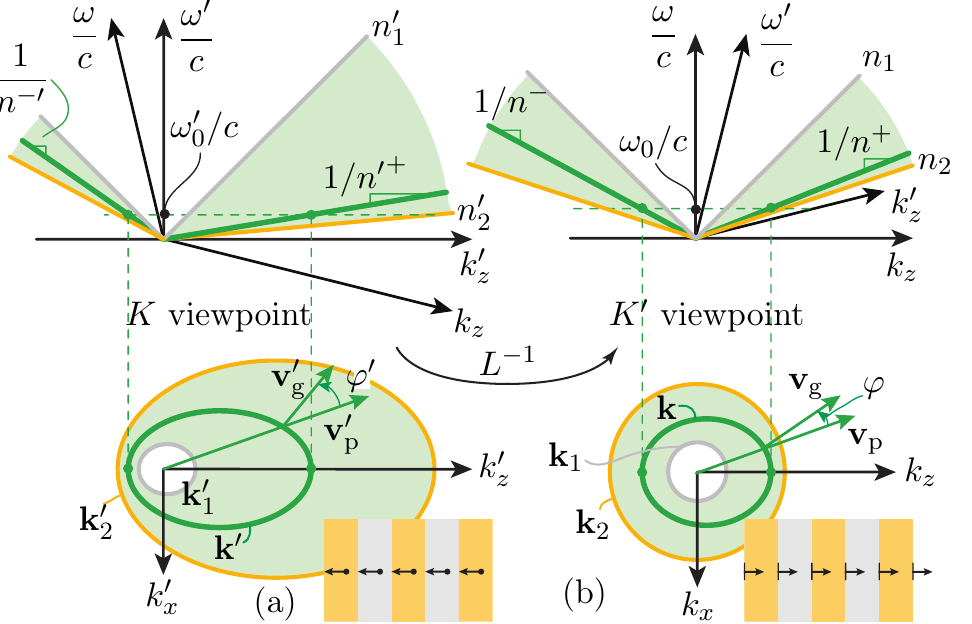}
        \caption{\small Spectral analysis for a 1+1D-$(z;t)$ USTEM metamaterial under oblique [$(k_z,k_x)$] incidence with modulation traveling in the $+z$ direction, including the $(\omega,k_z)^{(\prime)}$-cuts at $k_x^{(\prime)}=0$ of the $(\omega,k_z,k_x)^{(\prime)}$ cones (top panels) and $(\omega_0,k_z,k_x)^{(\prime)}$ isofrequency projections (bottom panels), compared with the cases of the bulk constituent media, with refractive indices $n_1'(\ve{k}')$ and $n_2'(\ve{k}')$ or $n_1$ and $n_2$. (a)~$K'$ viewpoint, with Fizeau-Fresnel drag, due to the motion of matter, in the $-z$ direction. (b)~$K$ viewpoint, with ``inverse drag'', due to the motion of the interfaces.}
        \label{fig:K_Kp_drag}
        \vspace{-4mm}
    \end{minipage}
    \hspace{0.01\textwidth}
    \begin{minipage}{0.3\textwidth}
        \includegraphics[width=1\textwidth]{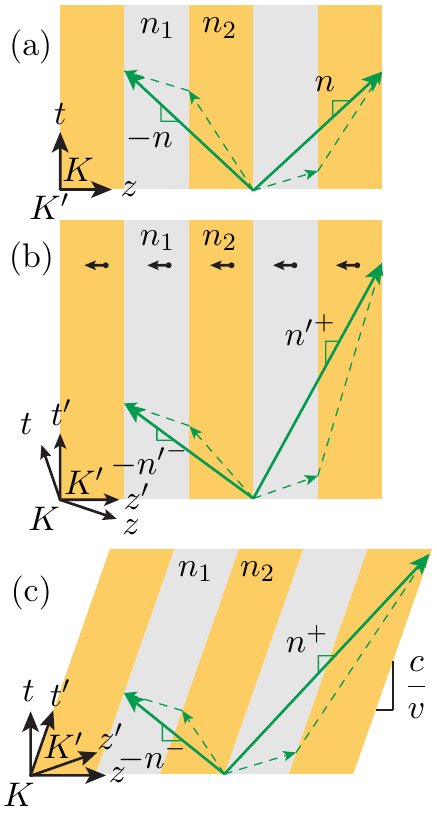}
        \caption{\small GSTEM weighted averaging deflection. (a)~SEM. (b)~USTEM, $K'$ viewpoint. (c)~USTEM, $K$ viewpoint.}
        \label{fig:GSTEM_weighting}
    \end{minipage}
\end{figure}

The fundamental properties of a GSTEM metamaterial, in the frame of interest ($K$), may be inferred from these preliminary considerations with the help of the spectral graphs provided in Fig.~\ref{fig:K_Kp_drag}, with Fig.~\ref{fig:K_Kp_drag}(a) representing the problem from the $K'$ viewpoint and Fig.~\ref{fig:K_Kp_drag}(b) from the $K$ viewpoint. Since the Fresnel-Fizeau drag occurs in the $-z'(=-z)$ direction in $K'$, as suggested in the panel at the extreme right bottom of the figure, the wave velocity, $v_{\mathrm{g}z}'=v_{\mathrm{p}z}'=c/n'$ (nondispersive medium assumption), is increased in the $-k_z'$ (or $-z'$) direction and decreased in the $+k_z'$ direction, i.e., $1/n^{-\prime}>1/n^{+\prime}$, so that the spectral cone is titled towards the positive $k_z'$ direction [top panel of Fig.~\ref{fig:K_Kp_drag}(a)]. As a result, the isofrequency curve at $\omega'=\omega_0'$ is a right-shifted ellipse [bottom panel of Fig.~\ref{fig:K_Kp_drag}(a)], located between the $n_1'(\ve{k}')$ and $n_2'(\ve{k}')$ also right-shifted ellipses, which is a manifestation of the expected bianisotropy (off-centered ellipse $\rightarrow$ bianisotropy, centered ellipse $\rightarrow$ anisotropy, centered circle $\rightarrow$ isotropy) (Supp. Mat.~\ref{sec:cryst_disp_rel_Kp}), and associated \emph{nonreciprocity}~\cite{Caloz_PRAp_10_2018}. The counterclockwise rotation, by the angle $\varphi'$, of the group velocity, $v_\mathrm{g}'(\ve{k}')=\nabla'_{\ve{k}'}\omega'(\ve{k}')$, with respect to the phase velocity, $v_\mathrm{p}' =(\omega'/k')\uve{k}'$, corresponds to the addition of a $-z'$ directed component to the velocity, and thus clearly shows the backward effect of the Fresnel-Fizeau drag.

We finally need to perform the required (Lorentz, Rindler, etc.) inverse transformation from $K'$ to $K$ to complete the resolution of the electrodynamics problem at hand. Obviously, the elimination of the motion of the atoms and molecules in the transition from $K'$ to $K$ eliminates motion-related bianisotropy, since $v_\mathrm{atom}=0$. In contrast, the exact nature (bianisotropic, anisotropic, biisotropic, homoisotropic) of the $K$ solution is difficult to predict and can be precisely found only by performing the exact $K'$-to-$K$ inverse transformation (inverse \emph{Lorentz} transformation in the USTEM case of Fig.~\ref{fig:K_Kp_drag}) (Supp. Mat.~\ref{sec:cryst_disp_rel_K}). Figure~\ref{fig:K_Kp_drag}(b) shows a typical result. Note that the $n_1'(\ve{k}')$ and $n_2'(\ve{k}')$ isofrequency ellipses have transformed to simple centered circles, corresponding to the expected curves for the assumed isotropic constituent media with (scalar) refractive indices $n_1$ and $n_2$. Interestingly, the USTEM curve is still a right-shifted ellipse, with a deflection of the group velocity by an angle of $\varphi$, still in the counterclockwise direction with respect to the phase velocity; this deflection is necessarily an effect of the moving-modulation \emph{interfaces}, since matter motion does not exist anymore. We have here $\varphi<\varphi'$, indicating that the deflection due to the modulation ($K$ frame) is smaller than that due to the matter ($K$ frame), but the effect is greater, and can reach $\varphi>\varphi'$, if the constitutive media have less-than-one indices ($n_1,n_2<1$) (plasma-type media). Most importantly, the observed \emph{light deflection} is contradirectional to the modulation ($+z$), and hence \emph{opposite to the direction of the drag for a moving body}, consistently with the finding in~\cite{Huidobro_PNAS_2019}.

This GSTEM deflection effect is quite distinct from the Fresnel-Fizeau drag. Indeed, it does not involve any motion of matter that would ``push'' or ``pull'' -- i.e., \emph{drag} -- light. It is rather an effect of \emph{spacetime weighted averaging}, as first suggested in \cite{Deck_APH_10_2019}. This effect is illustrated Fig.~\ref{fig:GSTEM_weighting}. In the SEM problem, represented in Fig.~\ref{fig:GSTEM_weighting}(a), light spends on average the same amount of time in medium~1 and in medium~2 in the forward and backward directions, so that the corresponding effective metamaterial indices are equal, i.e., $n^+=n^-$. In the $K'$-USTEM problem, represented in Fig.~\ref{fig:GSTEM_weighting}(b), light is subjected to the conventional Fresnel-Fizeau drag, and propagates therefore faster in the backward (downstream) direction than in the forward (upstream) direction, so that $n^{+\prime}>n^{-\prime}$. In the $K$-USTEM problem, represented in Fig.~\ref{fig:GSTEM_weighting}(c), the spacetime slopes are the same as in the SEM problem, since no matter motion occurs, but the medium is spacetime-wise oblique, here tilting in the forward direction, and the following -- rather subtle! -- effect occurs due to this tilting. Consider, for simplicity, that $n_2\gg{}n_1$, as in the figure [where the vertical axes have been denormalized to ensure a restricted graphical aspect ratio without introducing (unphysical) superluminal light curves]. Then, light spends much more time (see dashed lines) in medium~2, the slower medium, in the forward direction, where it propagates almost parallel to the medium trajectory, than in the backward direction, where the propagation is relatively more perpendicular to the medium trajectory, while the ratio of the forward to backward traveled distances increases in a much smaller ratio, as clearly apparent in the figure. As a result, $n^+>n^-$, and hence $v^+<v^-$, consistently with the observation in Fig.~\ref{fig:K_Kp_drag}. This is really, as announced, a \emph{spacetime weighted averaging} phenomenon; an exact mathematical formula for this phenomenon is provided in \cite{Deck_APH_10_2019} (Supp. Mat.~\ref{sec:ST_WA}).

\begin{figure*}[h!]
	\centering
	\includegraphics[width=0.9\textwidth]{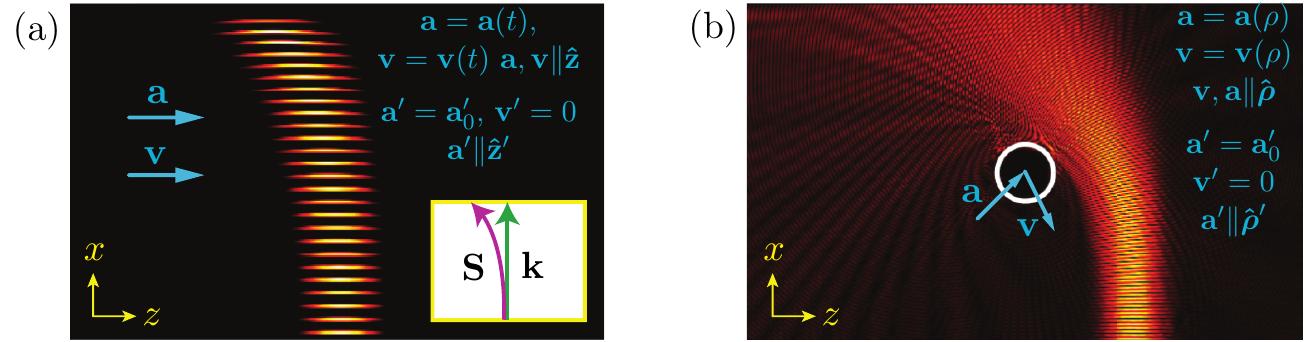}
\caption{\small Examples of ASTEM metamaterials and related transformations of a light beam (injected from the bottom). (a)~Rectilinear (Rindler metric~\cite{Moller_TR_1952,Rindler_PR_1960}) ASTEM, inducing curved light deflection. (b)~Black hole (Schwarzschild metric~\cite{Schwarzschild_BS_1916,MTW_2017}) ASTEM, attracting and partly absorbing light.}
\label{fig:ASTEM_illustr}
\end{figure*}

As ASTEM interfaces are the curved-spacetime generalization of USTEM interfaces (Sec.~\ref{sec:int_phys}), ASTEM metamaterials are the curved-spacetime generalization of USTEM metamaterials. Therefore, the principles exposed in conjunction with the USTEM graphs in Figs.~\ref{fig:K_Kp_drag} and~\ref{fig:GSTEM_weighting} largely extend to ASTEMs metamaterials, although their rigorous treatment requires a quantum jump from the \emph{theory of special relativity}~\cite{Einstein_1905,French_SR_1968}, routinely applied to USTEMs, to the \emph{theory of general relativity}~\cite{Einstein_1915,MTW_2017}. Figure~\ref{fig:ASTEM_illustr} presents two illustrative examples of ASTEM metamaterials~\cite{Bahrami_engrXiv_2022}. Figure~\ref{fig:ASTEM_illustr}(a) shows a rectilinear ASTEM metamaterial, which exhibits the modulation-contradirectional group delay deflection [$\ve{v}_\mathrm{g}(\ve{r})\|\ve{S}=\ve{E}\times\ve{H}$] and straight phase velocity [$\ve{v}_\mathrm{p}(\ve{r})\|\ve{k}$] propagation predicted in Fig.~\ref{fig:K_Kp_drag}; the beam curvature can be here qualitatively inferred from piecewise (straight) USTEM deflection. Figure~\ref{fig:ASTEM_illustr}(b) shows a rectilinear ASTEM black hole, which attracts and absorbs light like a cosmic black hole, an effect that is unattainable in simple graded-index lenses~\cite{Leonhardt_GL_2010}.
%


\section{Future Prospects}\label{sec:fut_prosp}
\vspace{-1mm}

Given their very fundamental nature and virtually unlimited diversity, GSTEMs have a formidable \emph{potential for scientific and technological innovation}. The \emph{scientific prospects} include 1)~the study of the properties of \emph{higher dimensional} (2+1D = 3D and 3+1D = 4D) GSTEM (unbounded) structures, 2)~the analysis of the scattering and diffraction at the interfaces and wedges of \emph{spacetime-truncated ~\cite{Deck_APH_10_2019} GSTEMs}, 3)~the exploration of \emph{new GSTEM geometries} (e.g., Rindler, Schwartzshild, Kerr, jerk, snap, crackle, etc.) in both the homogeneous and Bragg regimes, 4)~the extension of \emph{gravity analogs}, currently restricted to interface horizons~\cite{Faccio_AGF_2002}, such as white hole or Bose-Einstein condensate analogs, 5)~the elaboration of new electrodynamic computational tools for spacetime nonuniformities, using for instance foliation decomposition~\cite{Gourgoulhon_FGR_2012}, and for modulation-related multiphysics, and 6)~the investigation of novel GSTEM physics (e.g., superluminal and interluminal scattering, spacetime reversal, generalized spacetime metrics, spacetime quantum and sub-cycle phenomena). The \emph{technological prospects} include on the one hand the development of efficient modulation platforms and techniques (e.g., acoustic, electronic and optical) for the experimental implementation of the new GSTEM phenomena, and on the other hand the identification and demonstration of novel related applications. Many potential USTEM-related applications have been identified in~\cite{Caloz_TAP_PII_2019}; we expect that these applications will generalize to ASTEM-type systems, with extra opportunities offered by various spacetime curvatures and generalized spacetime ``chirping''.

\clearpage
{\small
\bibliographystyle{IEEEtran}
\bibliography{sample}}

\newpage
\appendix




\section{GSTEM Spacetime Diagrams}\label{sec:GSTEM_ST_diag}

The spacetime (or Minkowski) diagrams in Fig.~\ref{fig:GSTEMs_persp}, being $(z,ct)$ patterns, include the complete information on the GSTEM structures. A useful complement to these representations are the time-evolution ($t$) of corresponding purely spatial -- $(z,x)$ -- representations. Such representations are provided in the appended animation file \textbf{GSTEM\_Structures.mp4}.


\section{Fundamentals of Relativity}\label{sec:fund_rel}

A solid treatment of the electrodynamics of spacetime varying systems requires the utilization and adaptation of the principles and tools of the \emph{theory of relativity}, or \emph{relativity}. We shall therefore recall here the basic principles of this theory and establish the related essential tools.


\subsection{Basic Principles}\label{sec:rel_princ}

The \emph{fundamental principle of relativity} is that physical phenomena measured in spacetime reference frames that are moving with respect to each other are different from each other, due to the different spatial and temporal perspectives of the frames. In this statement, \emph{spacetime} is defined as a mathematical model that fuses the three dimensions of space and the one dimension of time into a single \emph{four-dimensional manifold}, which is itself a geometrical topological space that locally resembles the Euclidean space~\cite{Sher_HGT_2001}. Although not directly apparent in daily human experience, relativity effects play a major role in systems involving velocities approaching the speed of light, or \emph{relativistic} velocities, and they underpin myriads of aspects of modern science and technology~\mbox{\cite{MTW_2017,Vanbladel_RE_1984}}.

The theory of relativity, considered as a whole with Special Relativity (SR) and General Relativity (GR) combined, is based on the following four axioms:
\begin{enumerate}[I]
	\item \emph{First postulate} of SR: The laws of physics are identical in all non-accelerated, or  \emph{inertial}, systems~\cite{Einstein_1905,Pauli_TR_1958,French_SR_1968}.
	\item \emph{Second postulate} of SR: The speed of light in vacuum is the same -- $c=299,792.458$~m/s -- for all inertial systems, regardless of the motion of the light source~\cite{Einstein_1905,Pauli_TR_1958,French_SR_1968}.
	\item \emph{Equivalence principle} of GR: A uniform gravitational field is indistinguishable from a uniform acceleration field for a given reference frame~\cite{Einstein_1915,MTW_2017,Pauli_TR_1958,Weinberg_GC_1972}.
	\item \emph{Locality equivalence} between SR and GR: The laws of physics in a gravitational or accelerated field (GR) are locally identical to their SR-counterparts at every point of spacetime~\cite{Einstein_1915,MTW_2017,Pauli_TR_1958,Weinberg_GC_1972}.
\end{enumerate}
Axioms~I to~III follow from experimental evidence, while Axiom~IV is a consequence of the local flatness or differentiability of the \emph{spacetime} manifold~\cite{Tu_IM_2011}.

We shall deal here with the electrodynamics of both inertial -- and hence force-less -- and accelerated spacetime systems, which we shall hereafter refer to as \emph{uniform-velocity (UV)} (as USTEM) and \emph{nonuniform-velocity (NUV)} (as ASTEM) systems, respectively, in reference to an assumed inertial observation frame. UV and NUV systems respectively pertain to SR~\cite{Einstein_1905,Pauli_TR_1958,French_SR_1968} and GR~\cite{Einstein_1915,MTW_2017}. They are hence governed by Axioms~I and~II for the former case and by Axioms~III and~IV for the latter case. The first task in the study of the electrodynamics of such systems is to derive transformation relations between different reference frames for both the (direct and spectral) spacetime variables and the electromagnetic fields; this is accomplished in Sec.~\ref{sec:Lorentz_transf} for the UV case and in Sec.~\ref{sec:Rindler_transf} for the NUV case.

\begin{figure}[h!]
	\centering
	\includegraphics[width=0.7\textwidth]{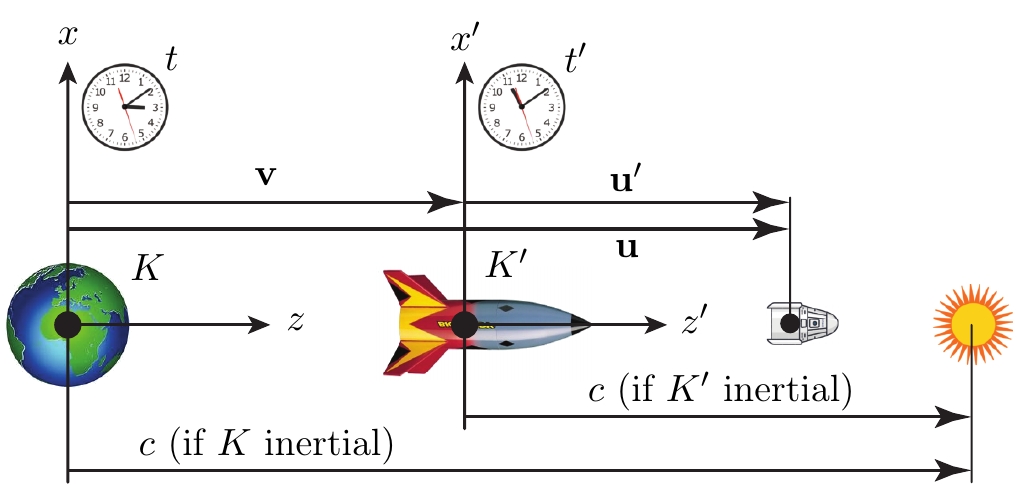}
	\caption{\small Relative reference frame setup used to describe spacetime systems.}
	\label{fig:relative_frames}
\end{figure}

The derivation of the variable and field transformations will be done with the help of the reference frame setup shown in Fig.~\ref{fig:relative_frames}. In this system, $K$ (here, Earth) is the frame where the experiment is conducted, and is hence called the \emph{laboratory frame}, while $K'$ (here, rocket) is a frame that moves at a velocity $\ve{v}=v\uve{z}$ relatively to $K$, and is hence called the \emph{moving frame}, or the \emph{rest frame} because comoving entities are at rest in that frame, despite the motion relatively to $K'$. The figure also shows a third element (here, spatial probe) in motion with velocities $\ve{u}$ and $\ve{u}'$ relatively to $K$ and $K'$, respectively, and propagating light, which is measured as having the same speed $c$ in $K$ and $K'$ if these frames are inertial, according to Axiom~II.


\subsection{Uniform-Velocity (UV) Transformations}\label{sec:Lorentz_transf}

Historically, the UV [$\ve{v}\neq\ve{v}(\ve{r},t)$ or $\ve{v}$-constant] transformation relations between the physical quantities in $K$ and $K'$ were first derived by Lorentz as the mathematical conditions for the form invariance of Maxwell's equations between inertial frames~\cite{Lorentz_1904}, and then shown by Einstein to follow directly from Axioms~I and~II~\cite{Einstein_1905}. The literature on the derivation of the these relations is rather obscure. We shall present here a clear and rigorous derivation, for both the spacetime variables and the electromagnetic fields, along with the velocity addition formula.

A priori, the relations between the spacetime variables $(z,t)$ of $K$ and $(z',t')$ of $K'$ (Fig.~\ref{fig:relative_frames}) could be anything. However, the homogeneity and isotropy, or the \emph{4D symmetry}, of spacetime that prevails in inertial systems demands that the relation between $K$ and $K'$ be the same at every point of spacetime. Therefore, 
\begin{subequations}\label{eq:relative_frames}
	\begin{equation}\label{eq:Lorentz_lin}
		\begin{Bmatrix}
			\frac{\partial}{\partial z}\left(\frac{\partial z'}{\partial z}\right)=0
			& \frac{\partial}{\partial t}\left(\frac{\partial z'}{\partial t}\right)=0 \\
			\frac{\partial}{\partial z}\left(\frac{\partial t'}{\partial z}\right)=0
			& \frac{\partial}{\partial t}\left(\frac{\partial t'}{\partial t}\right)=0
		\end{Bmatrix}\footnote{E.g., $\frac{\partial}{\partial z}\left(\frac{\partial t'}{\partial z}\right)=0$ because the relations between $t'$ and $z$ cannot depend on $z$ since space is assumed to everywhere be the same; similarly, $\frac{\partial}{\partial t}\left(\frac{\partial t'}{\partial t}\right)=0$ because the relation between $t'$ and $t$ cannot depend on $t$ since time is assumed to be always the same.},
	\end{equation}
	so that
	\begin{equation}\label{eq:Lorentz_formulas}
		\begin{Bmatrix}
			\frac{\partial z'}{\partial z}=A
			&\frac{\partial z'}{\partial t}=B \\
			\frac{\partial t'}{\partial z}=C
			& \frac{\partial t'}{\partial t}=D
		\end{Bmatrix},\;
		\text{or}\;
		\begin{Bmatrix}
			z'=Az+Bt \\
			t'=Cz+Dt
		\end{Bmatrix},
	\end{equation}
\end{subequations}
where $A$, $B$, $C$ and $D$ are constants. This indicates that the sought after relations must be \emph{linear}.

The constants $A$, $B$, $C$ and $D$ in Eqs.~\eqref{eq:Lorentz_formulas} may be found by successively enforcing the following physical conditions: 1)~motion of the origin of $K'$ ($z'=0$) at velocity $v$ relatively to $K$, 2)~motion of the origin of $K$ ($z=0$) at velocity $-v$ relatively to $K'$, 3)~uniformity of the speed of light, $c$ (Axiom~II), 4)~time reversal ($t,t'\rightarrow-t,-t'$) symmetry~\cite{Jackson_1998} upon successive direct (velocity $v$) and reverse (velocity $-v$) transformations, i.e., 
\begin{equation*}\label{eq:Lorentz_sys}
	\begin{Bmatrix}
		(z',z)=(0,vt)\Rightarrow A(vt)+Bt=0\Rightarrow Av+B\stackrel{1)}{=}0, \\
		(z,z')=(0,-vt')\Rightarrow-vt'=Bt,t'=Dt\Rightarrow B+vD\stackrel{2)}{=}0, \\
		\frac{z'}{t'}=c\rightarrow\frac{Az+Bt}{Cz+Dt}=c\stackrel{z=ct}{\Rightarrow}Ac+B-Cc^2-Dc\stackrel{3)}{=}0, \\
		\begin{pmatrix} A & -B \\ -C & D \end{pmatrix}
		\begin{pmatrix} A & B \\ C & D \end{pmatrix}
		=\begin{pmatrix} 1 & 0 \\ 0 & 1 \end{pmatrix}
		\Rightarrow
		A^2-BC\stackrel{4)}{=}1.
	\end{Bmatrix}
\end{equation*}

Solving this system of equations yields $A=D=\gamma$, $B=-\gamma v$ and $C=-\gamma\beta/c$, where
\begin{subequations}\label{eq:relative_frames}
	\begin{equation}\label{eq:gamma}
		\boxed{\gamma=\frac{1}{\sqrt{1-\beta^2}}},
		\quad\text{with}\quad
		\boxed{\beta=\frac{v}{c}},
	\end{equation}
	and whose substitution into the last relations of Eqs.~\eqref{eq:Lorentz_formulas} leads to the \emph{Lorentz transformation relations}
	\begin{equation}
		\boxed{\begin{Bmatrix}\label{eq:Lorentz_transf_rel}
			x'=x,\quad y'=y \\
			z'=\gamma(z-vt) \\
			ct'=\gamma\left(ct-\beta z\right)
		\end{Bmatrix}},
		\qquad\text{or}\qquad
		\boxed{\begin{Bmatrix}
			x=x',\quad y=y' \\
			z=\gamma(z'+vt') \\
			ct=\gamma\left(ct'+\beta z'\right)
		\end{Bmatrix}},
	\end{equation}
	where we have added trivial relations in the motion-less directions $x$ and $y$ for generalization to the observation direction $\ve{r}=x\uve{x}+y\uve{y}+z\uve{z}$.
\end{subequations}
Note that the direct and inverse relations [left and right in Eqs.~\eqref{eq:Lorentz_transf_rel}, resp.] are obtained from each other by exchanging the primed and unprimed quantities and changing the sign of $v$. Also note that $-(cdt)^2+dr^2=-(cdt')^2+dr^{\prime 2}$, where $r^{(\prime)2}=\sqrt{x^{(\prime)2}+y^{(\prime)2}+z^{(\prime)2}}$, which reveals that the quantity $ds^2=-d(ct)^2+r^2$, which is called the \emph{Minkowski metric}\footnote{The (UV) Minkowski metric has the signature $(-,+,+,+)$, which differs from the Euclidian signature $(+,+,+)$ by the extra and negative sign associated with time. In Sec.~\ref{sec:Rindler_transf}, we shall further encounter the (NUV) Rindler-Kottler-M{\o}ller transformation, whose signature is also $(-,+,+,+)$, but whose metric is distinct, due to acceleration.}, is an \emph{invariant of spacetime}.

\begin{figure}[h!]
	\centering
	\includegraphics[width=\textwidth]{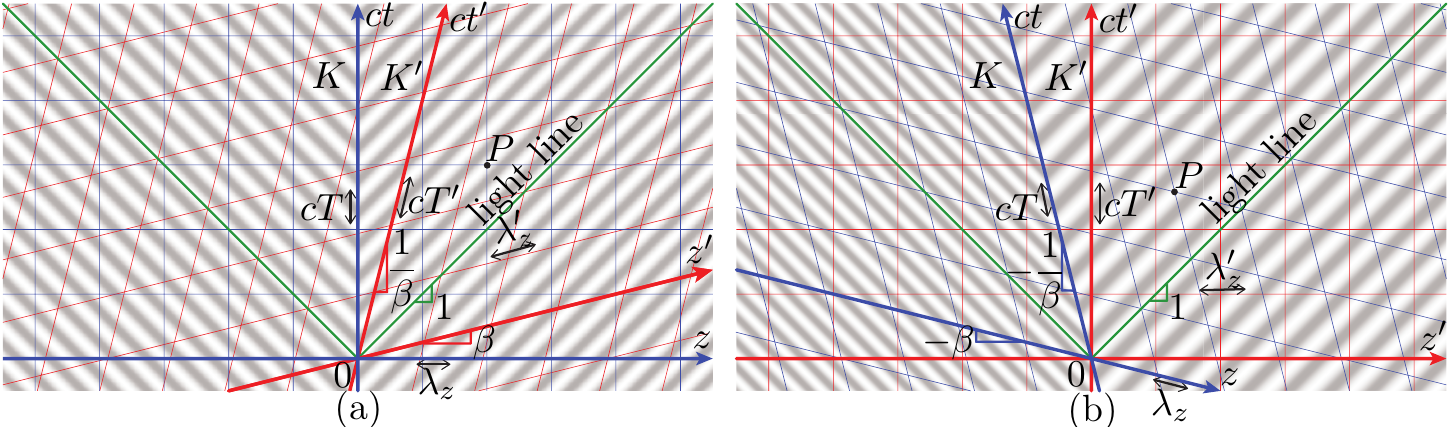}
	\caption{\small Superimposed $K$ (blue) and $K'$ (red) reference frames for $\ve{v}=v\ve{\hat{z}}$ (Fig.~\ref{fig:relative_frames}) with $v=\text{constant}$ (UV system), i.e., Lorentz frames, along with trajectory of light (green) and trajectories of the crests (white) and troughs (gray) of a monochromatic plane wave launched in free space to the right and left from the plane $z=0$. (a)~$K$ viewpoint. (b)~$K'$ viewpoint.}
	\label{fig:Lorentz_frames}
\end{figure}

Figure~\ref{fig:Lorentz_frames} plots the Lorentz reference frames corresponding to Eqs.~\eqref{eq:Lorentz_transf_rel}, from the $K$ viewpoint in Fig.~\ref{fig:Lorentz_frames}(a), where the axes $(z',ct')$ are obtained by setting $(ct',z')=(0,0)$ in the direct relations, and from the $K'$ viewpoint in Fig.~\ref{fig:Lorentz_frames}(b), where the axes $(z,ct)$ are obtained by setting $(ct,z)=(0,0)$ in the inverse relations. Projecting the intervals between successive crests or troughs of the monochromatic plane wave drawn in the background onto the axes $(z,z';ct,ct')$ immediately provides the corresponding wavelengths (or spatial periods) and (temporal) periods $(\lambda,\lambda';cT,cT')$. The differences in the two frames illustrate the Doppler effect, whereby the wave, which propagates in the $+z$-direction in the first quadrant, is seen with larger frequency ($\omega=2\pi/T$, blue shift) from the $-z$-direction, relatively counter-moving frame $K$, and with smaller frequency (red shift) from the $+z$-direction, relatively co-moving frame $K'$.

It is useful to also establish the spectral (Fourier-transform) counterparts of the spatial Lorentz relations~\eqref{eq:relative_frames}. These relations may be easily found from the latter by noting that $K$ and $K'$ observers agree on the phase of a monochromatic plane wave at any point $P$, with spacetime coordinates $(z,ct)\equiv(z',ct')$ (Fig.~\ref{fig:Lorentz_frames}), as corresponding to a crest, a trough, or any other level of the wave amplitude. Thus, we have $\phi'(z',ct')=\phi(z,ct)$, where $\phi=k_zz-\omega t$ and $\phi'=k_z'z'-\omega't'$ are the phase of the wave at $P$ in $K$ and $K'$, respectively, with $k_z^{(\prime)}=2\pi/\lambda_z^{(\prime)}$ and $\omega^{(\prime)}=2\pi/T^{(\prime)}$ being the corresponding wavenumbers (or spatial frequencies) and (temporal) frequencies, respectively. Substituting the inverse relations in Eq.~\eqref{eq:Lorentz_transf_rel} in the resulting relation, $k_zz-\omega t=k_z'z'-\omega't'$, factoring out $z'$ and $t'$, and noting that the last relation must hold for any $z'$ and $t'$, yields then
\begin{equation}\label{eq:spec_relative_frames}
	\boxed{\begin{Bmatrix}
		k_x'=k_x,\quad k_y'=k_y \\
		k_z'=\gamma\left(k_z-\frac{\beta}{c}\omega\right) \\
		\omega'=\gamma\left(\omega-vk_z\right)
	\end{Bmatrix}},
	\qquad\text{or}\qquad
	\boxed{\begin{Bmatrix}
		k_x=k_x',\quad k_y=k_y' \\
		k_z=\gamma\left(k_z'+\frac{\beta}{c}\omega'\right) \\
		\omega=\gamma\left(\omega'+vk_z'\right)
	\end{Bmatrix}},
\end{equation}
where the transformation between the direct and inverse relations is obtained, again, by exchanging the primed and unprimed quantities and by changing the sign of $v$. The last relations in Eq.~\eqref{eq:spec_relative_frames} provide the exact formulas for the Doppler frequency shift; specifically, the shifted frequencies are $\omega=\gamma\omega'(1\pm|\beta|\cos\theta')$ or $\omega'=\gamma\omega(1\mp|\beta|\cos\theta)$, where $\theta^{(\prime)}$ is the angle between the observation direction ($r^{(\prime)}$) and the motion direction ($z^{(\prime)}$), with the relative frequency shift being $\Delta\omega/\omega'=\pm\gamma|\beta|\cos\theta'\overset{|v|\ll c}{=}\pm|\beta|\cos\theta'$ or $\Delta\omega'/\omega=\mp\gamma|\beta|\cos\theta\overset{|v|\ll c}{=}\mp|\beta|\cos\theta$. Spectral spacetime diagrams, described in~\cite{Caloz_TAP_PII_2019}, provide a powerful graphical technique to determine both the temporal and the spatial frequency transitions induced by moving systems. 

The field transformation relations may now be simply found by applying Axiom~I to electrodynamics, viz., by enforcing the form invariance of Mawell's equations in $K$ and $K'$. This is accomplished by expressing the spacetime derivatives versus $K$ in terms of the spacetime derivatives versus $K'$, obtained from Eq.~\eqref{eq:Lorentz_transf_rel} as
\begin{equation}
	\begin{Bmatrix}
		\frac{\partial}{\partial x}=\frac{\partial}{\partial x'}\qquad\frac{\partial}{\partial y}=\frac{\partial}{\partial y'} \\
		\frac{\partial}{\partial z}=\frac{\partial}{\partial z'}\frac{\partial z'}{\partial z}+\frac{\partial}{\partial t'}\frac{\partial t'}{\partial z}
		=\gamma\left(\frac{\partial}{\partial z'}-\frac{\gamma\beta}{c}\frac{\partial}{\partial t'}\right) \\
		\frac{\partial}{\partial t}=\frac{\partial}{\partial z'}\frac{\partial z'}{\partial t}+\frac{\partial}{\partial t'}\frac{\partial t'}{\partial t}
		=\gamma\left(\frac{\partial}{\partial t'}-v\frac{\partial}{\partial z'}\right)
	\end{Bmatrix},
\end{equation}
inserting these relations into the $K$-frame Mawell's equations, and identifying with the $K'$-frame Mawell's equations. This gives the \emph{Lorentz field transformations}
\begin{equation}\label{eq:field_transf}
	\boxed{\begin{Bmatrix}
		E_x'=\gamma\left(E_x-vB_y\right) \\
		E_y'=\gamma\left(E_y+vB_x\right)  \\
		E_z'=E_z \\
		D_x'=\gamma\left(D_x-\frac{\beta}{c}H_y\right) \\
		D_y'=\gamma\left(D_y+\frac{\beta}{c}H_x\right)  \\
		D_z'=D_z
	\end{Bmatrix}},
	\qquad\text{or}\qquad
	\boxed{\begin{Bmatrix}
		B_x'=\gamma\left(B_x+\frac{\beta}{c}E_y\right) \\
		B_y'=\gamma\left(B_y-\frac{\beta}{c}E_x\right)  \\
		B_z'=B_z \\
		H_x'=\gamma\left(H_x+vD_y\right) \\
		H_y'=\gamma\left(H_y-vD_x\right)  \\
		H_z'=H_z
	\end{Bmatrix}},
\end{equation}
with the usual connection between the direct and the inverse relations.

Let us finally derive the \emph{velocity addition formula}. For this purpose, consider an object moving at the speed $u'=d z'/dt'$ in $K'$ (probe in Fig.~\ref{fig:relative_frames}). Taking the ratio of the differentials of the inverse Lorentz transformation formulas in Eqs.~\eqref{eq:Lorentz_transf_rel}, namely $dz=\gamma(dz'+vdt')$ and $dt=\gamma(dt'+\frac{\beta}{c}dz')$, gives the velocity of the object in $K$~\cite{Einstein_1905},
\begin{equation}\label{eq:v_addition}
	u=\frac{dz}{dt}=\frac{dz'+vdt'}{dt'+\frac{\beta}{c}dz'}=\frac{\frac{dz'}{dt'}+v}{1+\frac{\beta}{c}\frac{dz'}{dt'}}=\frac{u'+v}{1+\frac{\beta}{c}u'},
\end{equation}
which differs from the simple (Galilean) sum ($u'+v$) by the factor $\beta u'/c$ in the denominator. In the case of a relatively small $K'$-frame velocity ($\beta\ll 1$) and moderate $u'$ velocity, the denominator of the last expression in~\eqref{eq:v_addition} can be approximated by its first-order Taylor series, which, assuming an isotropic propagation medium with refractive index $n$ so that $u'=c/n$, simplifies the result to
\begin{equation}\label{eq:v_addition}
	u=(u'+v)\left(1-\frac{\beta}{c}u'\right)=\frac{c}{n}+v\left(1-\frac{1}{n^2}\right)-\frac{v^2}{cn}\approx\frac{c}{n}+v\left(1-\frac{1}{n^2}\right),
\end{equation}
the original (approximate) formula of Fizeau~\cite{Fizeau_1851}.

Generalizations of the formulas derived in this section to arbitrary vectorial velocities, $\ve{v}$, are available in many textbooks (E.g.,~\cite{Vanbladel_RE_1984,Kong_EWT_2008}).


\subsection{Rindler Transformations}\label{sec:Rindler_transf}

There is a virtually unlimited diversity of possible NUV [$\ve{v}=\ve{v}(\ve{r},t)$] systems, as may be realized by considering their numerous gravitational counterparts (Axiom~III) in cosmology, without even mentioning modulation-specific (e.g., equivalent negative mass or non-progressive motion) extensions. Examples include spacetime systems with spherical (Schwarzschild) symmetry, charged spherical (Reissner–Nordstr\"{o}m) structure, or charged rotationally-moving spherical (Kerr–Newman) gravitation or acceleration features, which are commonly studied in GR~\cite{MTW_2017,Weinberg_GC_1972,Carroll_SG_2019}. Here, we shall restrict our attention to NUV systems, $K'$, that involve \emph{rectilinear} (locally identical to spherical) motion relatively to an inertial observer, $K$, i.e., $\ve{v}=v(\ve{r},t)\uve{z}$, corresponding to the acceleration  $\ve{a}=(dv/dt)\uve{z}$ (Fig.~\ref{fig:relative_frames}). This restriction, although considerable, will fortunately not prevent us from capturing the essence of the phenomena occurring in NUV systems. 

In addition to the restriction of rectilinear motion, we shall assume that the acceleration of the considered NUV entities is independent of time in their reference frame\footnote{Time-wise constant acceleration implies ultimately exceeding the speed $c$. This is not at odds with Axiom~II since this axiom applies only to \emph{inertial} frames whereas $K'$ is \emph{noninertial}. The speed of light (and objects) in noninertial frames is not limited to $c$. For instance, in the counter-rotating direction of a vacuum-based Sagnac system~\cite{Sagnac_1_1913}, it exceeds~$c$ in the rotating frame. Here, $u$ is limited to $c$ because $K$ is inertial, but $u'$ is not since $K$ is noninertial.}, i.e., $a'\neq a'(t')$, while allowing it to vary in space, i.e., $a'=a'(z')$. This is equivalent to assuming that the accelerated entities \emph{experience} a force that may vary in space but that is time-wise constant any point of space, an assumption that is almost universally done in GR, both because no simple transformation relations could be found otherwise and because it corresponds to the physical reality of typical gravitational systems\footnote{The gravitational force of a celestial body might depend on the position of the observer around it, if its mass is nonuniformly distributed (e.g., moon), but it does not depend on the time of observation at any given position.}. This assumption is also reasonable for an STM system, whose acceleration may depend on the position, e.g., via a nonuniform background medium, but will typically not depend on time at a given position.

Most textbooks on GR (e.g., ~\cite{Weinberg_GC_1972,MTW_2017,Carroll_SG_2019}) derive the NUV transformation relations from metric tensors associated with differential geometry. This approach, while being the most general and the most powerful existing one, offers unfortunately little insight into the physics of NUV systems. To avoid this inconvenience, and to maintain the self-consistency spirit of this appendix, we shall present here an intuitive, yet rigorous, derivation of the NUV transformation relations, as in the case of uniform velocity.

The Lorentz transformations [Eqs.~\eqref{eq:relative_frames}] are linear because of the symmetry of spacetime that prevails in inertial systems. Such symmetry is broken in noninertial systems, as may be realized by thinking for instance of the gravitational potential of a massive object, which decays as $1/r$ from its center, from a high level at its surface to virtually zero farther in deep space. Therefore, the NUV relations between $K$ and $K'$ must necessarily be \emph{nonlinear}. Consequently, we must adopt a more elaborate derivation strategy than in the UV case.

\begin{figure}[h!]
	\centering
	\includegraphics[width=\columnwidth]{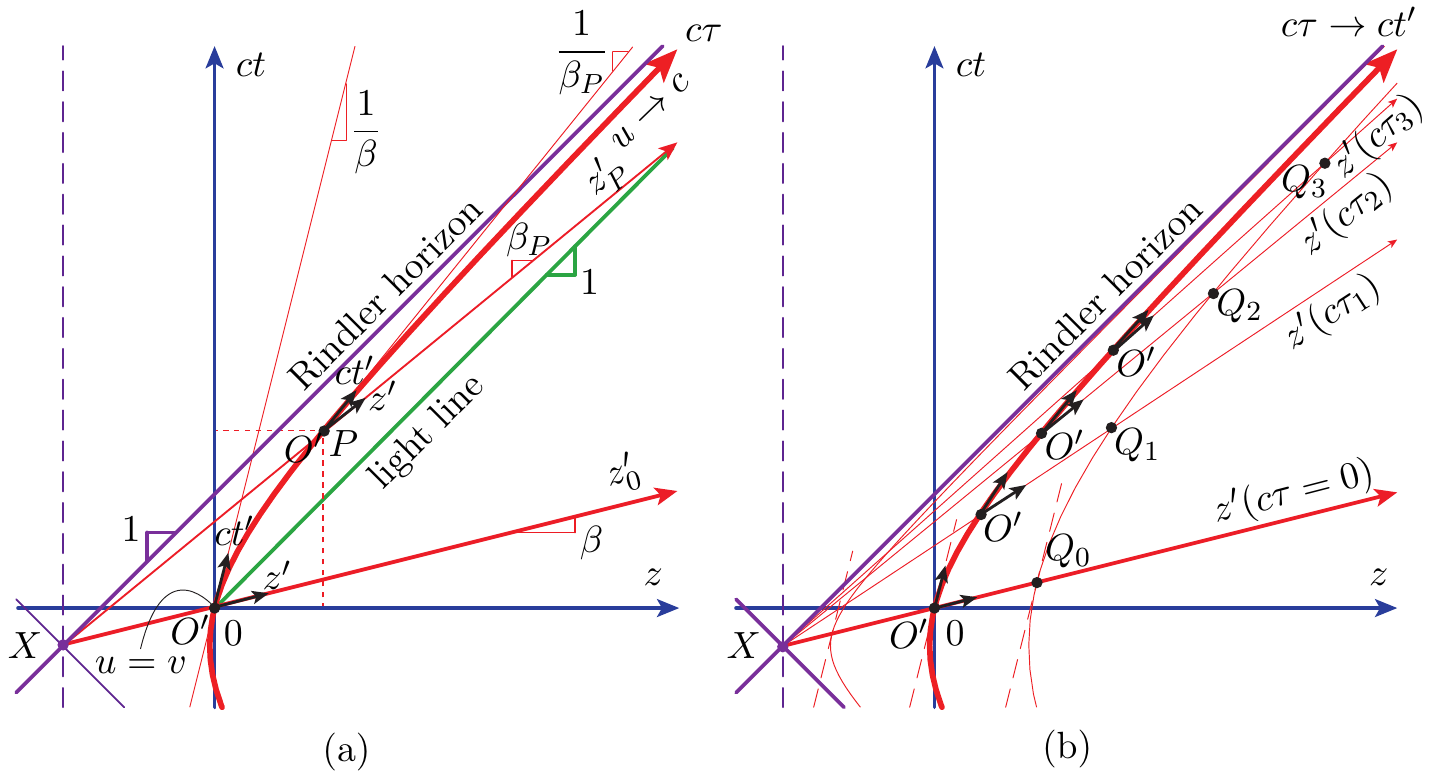}
	\caption{\small Graphical aid for the derivation of the NUV transformation relations. (a)~Local Lorentz frames $(O',z',ct')$ at the origin of $K$ and at point $P$. (b)~Formation of the Rindler-Kottler-M{\o}ller coordinate system with the local frame moved along the $c\tau$ curve, which ultimately becomes the $ct'$ axis.}
	\label{fig:KM_constr}
\end{figure}

We shall determine the NUV transformation relations by invoking Axiom~IV, viz., by applying the UV transformation relations [Eqs.~\eqref{eq:Lorentz_transf_rel}] locally in the NUV system of interest. This will be done with the help of Fig.~\ref{fig:KM_constr}. As we have seen in Fig.~\ref{fig:Lorentz_frames}(a), the UV coordinate axis $ct'$ corresponds to the trajectory of a UV object that is co-moving with $K'$ in $K$. Similarly, Fig.~\ref{fig:KM_constr}(a) plots the trajectory of an \emph{accelerated} object (e.g., spatial probe in Fig.~\ref{fig:relative_frames}) in $K$, $c\tau$, which will ultimately become the NUV $ct'$ axis upon application of the differential forms of the UV relations for locality. This trajectory has a curvature, which corresponds to the assumed constant acceleration (and hence constant force) in $K'$, $a_0'$. Consequently, the object will accelerate along its trajectory with acceleration $a(t)$ until it reaches the speed of light, $u(t)\rightarrow c$, in $K$, as indicated in the figure.

According to Axiom~IV, the laws of SR (Sec.~\ref{sec:Lorentz_transf}) may be applied \emph{locally} to the GR problem at hand. Specifically, they may be applied at the origin of $K$ in Fig.~\ref{fig:KM_constr}(a), where the accelerated (NUV) frame, $(z',ct')$, locally reduces to the Lorentz (UV) frame $K'$ $(O';z',ct')$ shown in the figure. At that point, we have $u=v$ (same velocity for the probe as for the rocket relatively to $K$), so that $u'=0$ (Fig.~\ref{fig:relative_frames}). Taking the derivative of the SR velocity addition formula~\eqref{eq:v_addition} with respect to $t$ leads then to the local relation (Sec.~\ref{sec:der_dudt_vs_dupdtp})
\begin{equation}\label{eq:a_ap_rel}
a=\frac{du}{dt}=\frac{a_0'}{\check{\gamma}^3},
\quad\text{where}\quad\check{\gamma}=\frac{1}{\sqrt{1-\left(\frac{u}{c}\right)^2}},
\end{equation}
between the accelerations in $K$ ($a$) and in $K'$ ($a_0'$) at the origin.

Integrating the relation~\eqref{eq:a_ap_rel} from $0$ at an arbitrary point $P$ on the trajectory $c\tau$, specifically integrating over the velocity from $v$ to $u$, over time from $0$ to $t$ and over space from $0$ to $z$, delocalizes the problem and leads to the global velocity and position functions along the entire trajectory $c\tau(z,t)$ in $K$ (Sec.~\ref{sec:der_nu_uz_vs_t}),
\begin{subequations}\label{eq:nu_uz_vs_t}
	\begin{equation}\label{eq:nu_u_vs_t}
	u(t)=\frac{a_0't+\gamma v}{\sqrt{1+\left(\frac{a_0't}{c}+\gamma\beta\right)^2}}
	\end{equation}
	and
	\begin{equation}\label{eq:nu_z_vs_t}
	z(t)=\frac{c^2}{a_0'}\left[\sqrt{1+\left(\frac{a_0't}{c}+\gamma\beta\right)^2}-\gamma\right].
	\end{equation}
\end{subequations}
The shape of the trajectory corresponding to Eq.~\eqref{eq:nu_z_vs_t} may be better visualized upon moving the $t$-dependent right-hand side square root to the left-hand side and squaring the resulting equation. This yields
\begin{equation}\label{eq:nu_K_hyperb}
\left(z+\frac{c^2}{a_0'}\gamma\right)^2-\left(ct+\frac{c^2}{a_0'}\gamma\beta\right)^2
=\left(\frac{c^2}{a_0'}\right)^2,
\end{equation}
which is the equation of a hyperbola centered at the point $X\equiv(-c^2\gamma/a_0',-c^2\gamma\beta/a_0')$ and with asymptotes $(ct)=\pm(z+c^2\gamma/a_0')-c^2\gamma\beta/a_0'$, in $K$, as shown in Fig.~\ref{fig:KM_constr}(a), where only the right branch of the hyperbola is considered given the assumption of motion in the $+z$-direction.

We can now globalize the problem from the hyperbola $c\tau$ to the entire spacetime plane. For this purpose, we consider a UV local frame $(O';z',ct')$ that moves along the trajectory $c\tau$, as indicated in Fig.~\ref{fig:KM_constr}(b), and describe the points $Q(c\tau)$, noted $Q_n$ in the figure\footnote{The points $Q_n$ in the figure are placed on constant $z'$ curve to emphasize the corresponding coordinate, but they may be of course placed on \emph{any} of such coordinate curve to eventually cover the entire (continuous) spacetime plane.}, in the spatial neighborhood of $O'$ by the coordinates \mbox{$(z',ct'=0)$} in that local frame. We use then a $c\tau$-parametrized version of the local Lorentz relations [Eqs.~\eqref{eq:Lorentz_transf_rel}], $z(z',ct')=\check{\gamma}(\tau)(z'+vt')$ and $ct(z',ct')=\check{\gamma}(\tau)(ct'+\check{\beta}(\tau)z')$, where $\check{\gamma}(\tau)$ and $\check{\beta}(\tau)$ are the $\tau$-parametrized versions of $\gamma$ and $\beta$ in Eqs.~\eqref{eq:gamma} ($\tau=0$) , with $t'=0$, and find the following form for the $K$-frame coordinates \mbox{of $Q$}:
\begin{equation}\label{eq:KM_offset}
	z=z_{O'}(c\tau)+\check{\gamma}(c\tau)z',\quad
	ct=ct_{O'}(c\tau)+\check{\gamma}(c\tau)\check{\beta}(c\tau)z',
\end{equation}
where $z_{O'}(c\tau)$ and $ct_{O'}(c\tau)$ are the coordinates of $O'$ and $z'$ is the distance of $Q$ to $O'$, as apparent in the figure. The $c\tau$ parametrization is performed, invoking again Axiom~IV, by first integrating the local relation $dt=\check{\gamma}(t)dt'$ with $v=u(t)$ over time from $0$ to $t$ for $t$ and from $0$ to $\tau$ for $t'$ in the left- and right-hand sides, respectively, which yields the $t=\tau$ mapping relation (Supp. Mat.~\ref{sec:der_loc_KM_transf})
\begin{equation}\label{eq:t_vs_tp_KM}
t(\tau)=\frac{c}{a_0'}\left\{\sinh\left[\frac{a_0'\tau}{c}+\sinh^{-1}\left(\gamma\beta\right)\right]-\gamma\beta\right\},
\end{equation}
and inserting then this relation into the $c\tau$-trajectory velocity formula~\eqref{eq:nu_u_vs_t} to obtain the functions $\check{\gamma}(c\tau)$ and $\check{\beta}(c\tau)$. From that point, $z_{O'}(c\tau)$ and $ct_{O'}(c\tau)$ are obtained by substituting these functions into Eqs.~\eqref{eq:nu_z_vs_t} and~\eqref{eq:t_vs_tp_KM}, and Eqs.~\eqref{eq:KM_offset} take the explicit form (Supp. Mat.~\ref{sec:der_KM_transf_app})
\begin{subequations}\label{eq:KM_transf}
	\begin{equation}\label{eq:KM_transf_indir}
		\boxed{\begin{Bmatrix}
			x=x',\quad y=y' \\
			z=(z'+z_0)\cosh(ct'/z_0+\xi)-z_a \\
			ct=(z'+z_0)\sinh(ct'/z_0+\xi)-z_b
		\end{Bmatrix}},
	\end{equation}
	\begin{equation}\label{eq:KM_transf_dir}
			\text{or}\quad
			\boxed{\begin{Bmatrix}
			x'=x,\quad y'=y \\
			z'=\sqrt{(z+z_a)^2-(ct+z_b)^2}-z_0 \\
			ct'=z_0\left[\tanh^{-1}\left(\frac{ct+z_b}{z+z_a}\right)-\xi\right]
		\end{Bmatrix}},
	\end{equation}	
\end{subequations}
where $z_0=c^2/a_0'$, $z_a=\gamma z_0$, $z_b=\gamma\beta z_0$ and $\xi=\sinh^{-1}(\gamma\beta)$, and where $\gamma$ and $\beta$ are still given by Eq.~\eqref{eq:gamma}. Equations~\eqref{eq:KM_transf} are the \emph{Rindler-Kottler-M{\o}ller transformation relations}~\cite{Moller_TR_1952}\footnote{In fact, one distinguishes the Rindler and Kottler-M{\o}ller relations, where the latter are a shifted version of the former relations~\cite{Rindler_PR_1960} in $K$. The Rindler relations have the advantage of being simpler and symmetric with respect to the axes of $K$. However, they are not congruent with the Lorentz relations -- their $K'$-origin does not coincide with the origin of $K$ -- which prevents direct comparisons with the Lorentz relations.}, which are the sought after NUV counterparts of the Lorentz relations~\eqref{eq:relative_frames}.

The following comments are in order about Fig.~\ref{fig:KM_constr}(b) and Eqs.~\eqref{eq:KM_transf}. First, each hyperbolic coordinate curve corresponds to a time evolution (or aging) that is specifically experienced by a hypothetical object following it; therefore, the different times experienced along the different hyperbolas are called the \emph{proper times}, with clocks ticking slower and faster on the left and right of the $ct'$ axis, respectively. Second, each oblique straight coordinate line ($z'$) corresponds to locations where all the points of $K'$ are synchronized; these lines are therefore called \emph{simultaneity planes}. Third, the $z'=\text{constant}$ coordinate lines are the right branches of the hyperbolas described by Eq.~\eqref{eq:nu_K_hyperb} with $a_0'$ replaced by the parameter $a'$ in the right-hand side of the relation and having the vertices $(-c^2\gamma/a_0'+c^2/a'^2,-c^2\gamma\beta/a_0')$ (same center and asymptotes but different vertices); inserting the inverse Lorentz relations into the new equation, setting $t'=0$ (time-invariant acceleration) gives the $z'$-dependent acceleration
\begin{equation}
	a'(z')=\frac{c^2}{z'+c^2/a_0'},
\end{equation}
where $a'(0)=a_0'$, as expected. Finally, we have $-(cdt)^2+dr^2=-[(z+z_0)\zeta]^2(cdt')^2+dr'^2$ with $\zeta=z/z_0$ (Sec.~\ref{sec:KM_metric_sm}), where $ds'^2=-[(z+z_0)\zeta]^2(cdt')^2+dr'^2$ is the \emph{Kottler-M{\o}ller metric}, corresponding to the Minkowski metric $ds^2=-(cdt)^2+dr^2$ already encountered in Sec.~\ref{sec:Lorentz_transf}).

\begin{figure}[h!]
	\centering
	\includegraphics[width=\columnwidth]{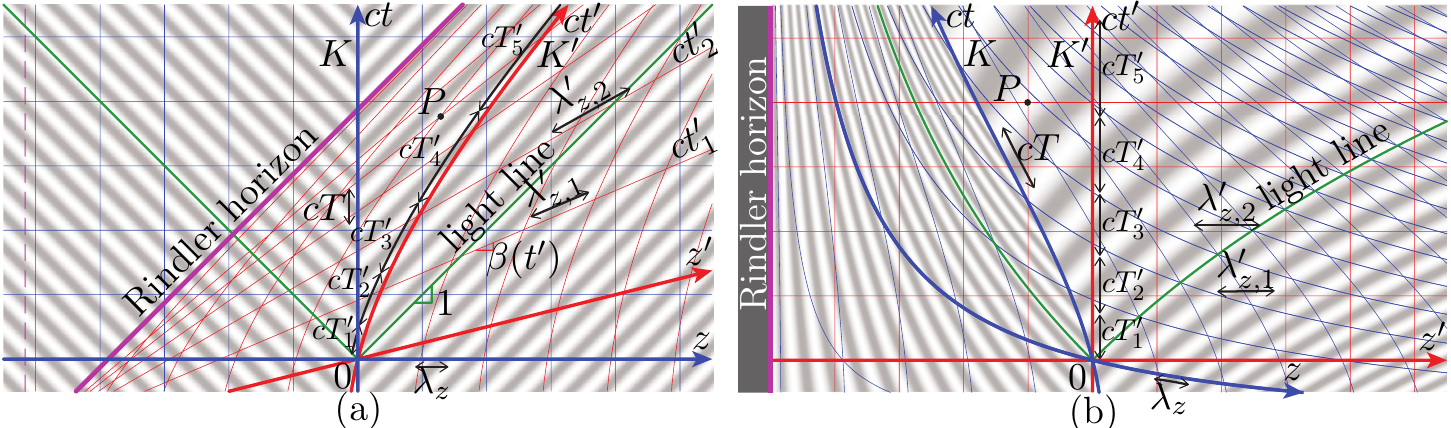}
	\caption{\small Superimposed $K$ (blue) and $K'$ (red)	reference frames for $\ve{v}=v\ve{\hat{z}}$ (Fig.~\ref{fig:relative_frames}) with $v=v(t)$ (NUV system) and $a'=da'/dt'=a_0'=\text{constant}$ (uniform proper acceleration), i.e., Lorentz -- Rindler-Kottler-M{\o}ller frames, along with trajectories of crests (white) and troughs (gray) of a monochromatic plane wave launched in free space from $K$ to the right and left from the plane $z=0$. (a)~$K$ viewpoint. (b)~$K'$ viewpoint.}
	\label{fig:KM_frames}
\end{figure}

Figure~\ref{fig:KM_frames} shows the corresponding frame-wave representation, as the NUV counterpart of the UV representation in Fig.~\ref{fig:Lorentz_frames}, from the $K$ viewpoint in Fig.~\ref{fig:KM_frames}(a) and from the $K'$ viewpoint in Fig.~\ref{fig:KM_frames}(b). The \emph{curvature of spacetime}, specific to GR~\cite{Einstein_1915,MTW_2017,Pauli_TR_1958,Weinberg_GC_1972}, is immediately apparent in these graphs. The figure also shows some wavelengths (or spatial periods) and (temporal) periods $(\lambda,\lambda';cT,cT')$. In contrast to those of the UV case (Fig.~\ref{fig:Lorentz_frames}), these quantities vary in spacetime, corresponding to the transformation of a monochromatic and monodirectional wave into a spacetime frequency rainbow\footnote{The temporal frequency rainbow is the chirping effect~\cite{Saleh_Teich_FP_2019}, which may be seen as an extension of the Doppler effect, while the the spatial frequency rainbow is akin to diffraction.}. Particularly interesting is the apparition of an \emph{event horizon}, called the \emph{Rindler horizon} for the present case of rectilinear, constant proper acceleration, which is a boundary of spacetime that can be crossed only one way, as the boundary of a black hole~\cite{MTW_2017,Weinberg_GC_1972}.

Given the aforementioned rainbow transformations, simple spectral relations such as the UV ones in Eqs.~\eqref{eq:spec_relative_frames} do not exist. In contrast, closed-form expressions for the field transformation counterparts of Eqs.~\eqref{eq:field_transf} exist, and they may be easily derived, invoking again Axiom~IV by locally applying the Lorentz field relations~\eqref{eq:field_transf}. Consider for instance the first direct Lorentz relation in Eqs.~\eqref{eq:field_transf}, which becomes, upon parametrizing $\gamma$ as $\gamma(t)=\partial t/\partial t'$ (boost) and $v$ as $v(t)=\partial z/\partial t$, $E_x'=(\partial t/\partial t')E_x-(\partial t/\partial t')(\partial z/\partial t)B_y=(\partial t/\partial t')E_x-(\partial z/\partial t')B_y$. In these relations, the derivatives are found from the Rindler-Kottler-M{\o}ller relations~\eqref{eq:KM_transf} as $\partial t/\partial t'=(z'+z_0)\zeta\cosh(\zeta t'+\xi)/c$ and $\partial z/\partial t'=(z'+z_0)\zeta\sinh(\zeta t'+\xi)$. The corresponding field transformation relation is then
\begin{subequations}
\begin{equation}\label{eq:Exp_KM}
E_x'=(z'+z_0)\zeta\left[\cosh(\zeta t'+\xi)E_x/c-\sinh(\zeta t'+\xi)B_y\right],
\end{equation}
which, upon noting that $(z'+z_0)\zeta/c=(z'+z_0)(c/z_0)/c=(z'/z_0+1)=(1+a_0'z'/c^2)$, factoring out $\cosh(\zeta t'+\xi)$, and further noting that $\beta=v/c=(dz/dt)/c=\tanh(\zeta t'+\xi)$ [using the differentials the $dz$ and $cdt$ following from~\eqref{eq:KM_transf_indir}] and thus $\gamma=1/\sqrt{1-\beta^2}=\cosh(\zeta t'+\xi)$, may also be written as
\begin{equation}\label{eq:Exp_KM}
E_x'=(1+a_0'z'/c^2)\gamma\left(E_x-vB_y\right),
\end{equation}
\end{subequations}
The other relations may be found in a similar fashion. The results are, using the GR-metric notation
\begin{subequations}
    \begin{equation}
        \boxed{g_{00}=(1+a_0'z'/c^2)^2},
    \end{equation}
\begin{equation}\label{eq:field_transf}
	\boxed{\begin{Bmatrix}
		E_x'=\sqrt{g_{00}}\gamma\left(E_x-vB_y\right) \\
		E_y'=\sqrt{g_{00}}\gamma\left(E_y+vB_x\right)  \\
		E_z'=\sqrt{g_{00}}E_z \\
		D_x'=\gamma\left(D_x-\frac{\beta}{c}H_y\right) \\
		D_y'=\gamma\left(D_y+\frac{\beta}{c}H_x\right)  \\
		D_z'=D_z
	\end{Bmatrix}},
	\qquad\text{or}\qquad
	\boxed{\begin{Bmatrix}
		B_x'=\gamma\left(B_x+\frac{\beta}{c}E_y\right) \\
		B_y'=\gamma\left(B_y-\frac{\beta}{c}E_x\right)  \\
		B_z'=B_z \\
		H_x'=\sqrt{g_{00}}\gamma\left(H_x+vD_y\right) \\
		H_y'=\sqrt{g_{00}}\gamma\left(H_y-vD_x\right)  \\
		H_z'=\sqrt{g_{00}}H_z
	\end{Bmatrix}}.
\end{equation}
\end{subequations}

In the case of an $x$-polarized transverse electromagnetic (TEM) wave propagating in free space, where $B_y=E_x/c$, and in the limit $t'\rightarrow\infty$, the relation~\eqref{eq:Exp_KM} reduces to
\begin{equation}
E_x'
=(z'+z_0)(\zeta/c)\textrm{e}^{-(\zeta t'+\xi)}E_x.
\end{equation}
This relation reveals that, for a given $E_x$ measured in $K$, $E_x'$ increases (linearly) with $z'$ and decreases (exponentially) with $t'$. These double effect may be explained from the Doppler effect, whereby the electromagnetic field compresses in spacetime and hence increases in magnitude with decreasing co-moving relative velocity between the source and the observer, which occurs for increasing $z'$ and decreasing $t'$, as clearly visible in Fig.~\ref{fig:KM_frames}(b).


\section{Scattering Angles at a USTEM Interface under Oblique Incidence}\label{sec:obl_inc}

\begin{figure*}[h!]
	\centering
	\includegraphics[width=0.9\textwidth]{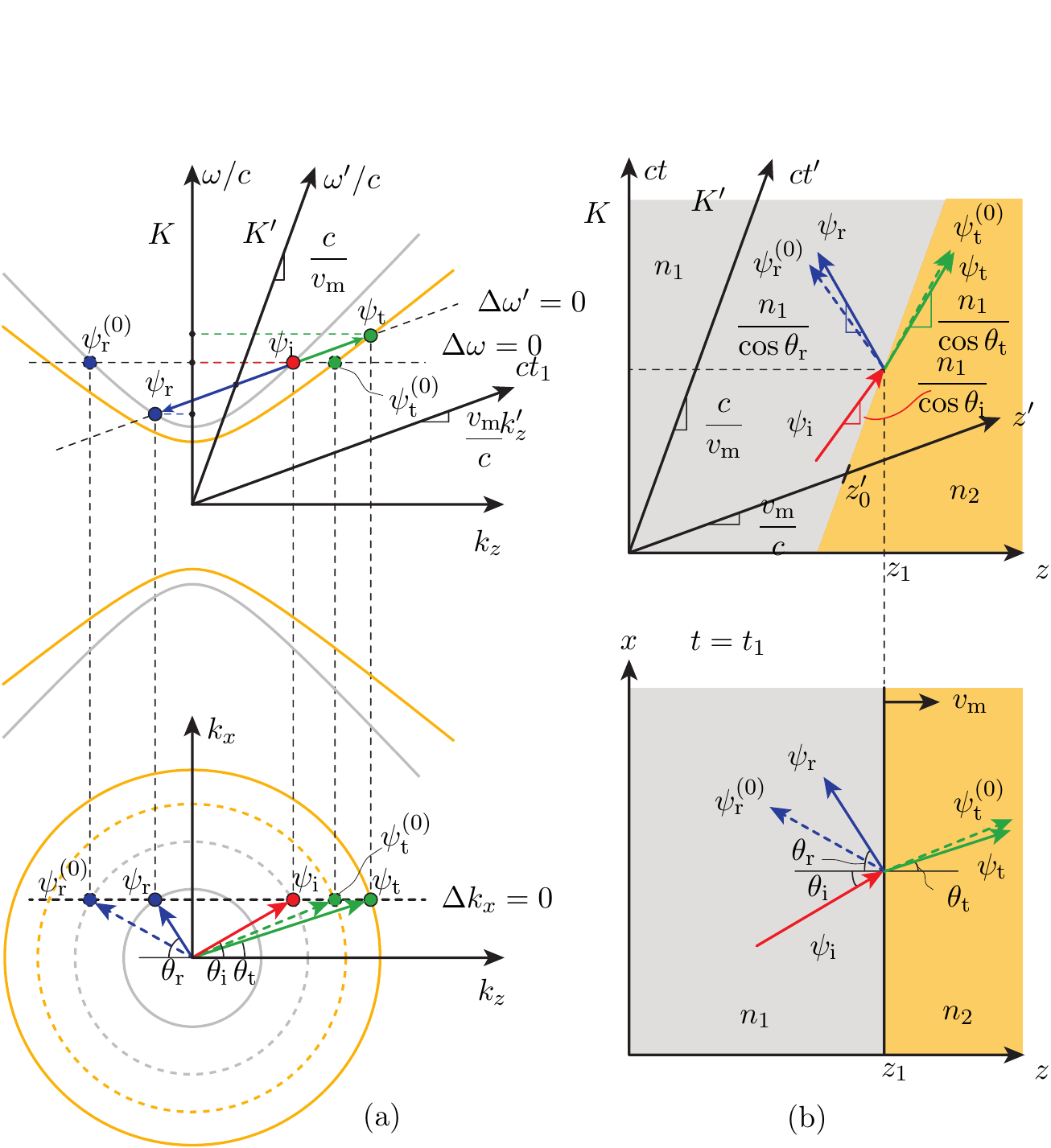}
	\caption{Electrodynamics of USTEM interfaces [Fig.~\ref{fig:GSTEMs_persp}(c)] for the case of oblique incidence [generalization of the normal-incidence case in Fig.~\ref{fig:STEM_interf}(a)] in the subluminal regime. The red, green and blue arrows correspond to the incident, reflected and transmitted waves, and the dashed arrows correspond to the spatial interface, where $v=0$. (a)~Inverse spacetime. (b)~Direct spacetime.}
	\label{fig:interf_obl_sub}
\end{figure*}

Figure~\ref{fig:interf_obl_sub} provides a graphical method to determine the scattering angles at a USTEM interface for oblique incidence in the subluminal regime. Figure~\ref{fig:interf_obl_sub}(a) shows two cuts of the dispersion diagram: the $\omega/c-k_z$ plane (top), corresponding to the dispersion diagram, and the $k_x-k_z$ plane (bottom), corresponding to the wavevector diagram. The dispersion diagram is obtained by the following steps:
\begin{enumerate}\vspace{-2mm}
    \item  plot the hyperbolas corresponding to $\omega^2n_{1,2}^2/c^2-k_z^2=k_{x\mathrm{i}}^2$, where $k_{x\mathrm{i}}$ is the transverse ($k_x$) component of the incident wave wavevector;
    \item plot the axes of the laboratory $K$ frame and the primed $K'$ frame;
    \item locate the incident wave spectrum coordinates $\omega_\mathrm{i}, k_{z\mathrm{i}}$;
    \item apply the conservation of frequencies in the primed frame, $\Delta \omega'=0$, to deduce the $\omega, k_{z}$ of the reflected and transmitted wave.
    \end{enumerate}\vspace{-2mm}
This is a generalization of the frequency transition plot of Fig.~\ref{fig:STEM_interf}(a) (right), with the only difference being that the dispersion curves are now hyperbolas instead of straight lines. 

The wavevector diagram is obtained by the following steps:
\begin{enumerate}\vspace{-2mm}
    \item draw the wavevector of the incident wave, which has a known angle;
    \item project the results of the dispersion diagram onto the wavevector diagram;
    \item apply the continuity of $k_x$ to deduce the scattering angles of the reflected and transmitted waves; \item  add the isofrequency circles corresponding to $k_x^2+k_z^2=\omega^2n^2/c^2$ (optional, added for completeness).
\end{enumerate}\vspace{-2mm}

Following this approach for a stationary interface and a modulated interface moving to the right ($+z$), as in Fig.~\ref{fig:interf_obl_sub}(a), we see that modulation to the right leads to a clockwise rotation of the reflected and the transmitted waves.

Figure~\ref{fig:interf_obl_sub}(b) presents the results in the direct space-time, by representing two cuts of the space-time trajectory of the waves: the $z-ct$ plane (top) and the $x-z$ plane (bottom). The angles in the $x-z$ plane are those from the bottom of Fig.~\ref{fig:interf_obl_sub}(a), and the slopes at the top are calculated as $n_z=c/v_z=c/(v\cos\theta)=n/\cos\theta$. This is a generalization of the space-time scattering of normal incidence, Fig.~\ref{fig:STEM_interf}(a) (left), to oblique incidence.

Similar graphs can be found in~\cite{Deck_PRB_2018} where only the reflected wave is derived. 

The deflection angles can be derived by inserting the Lorentz transformations of the wavenumber and the frequency~\eqref{eq:spec_relative_frames} into the phase-matching conditions, $k_{x\mathrm{i}}'=k_{x\mathrm{r}}'=k_{x\mathrm{t}}'$ and $\omega_\mathrm{i}'=\omega_\mathrm{r}'=\omega_\mathrm{t}'$. After some manipulations, we find then
\begin{subequations}\label{sub angular relation}
\begin{equation} 
     \cos{\theta_\mathrm{r}} = \frac{ (\beta^2 n_1^2+1) \cos{\theta_\mathrm{i}} - 2 \beta n_1 }{ \beta^2 n_1^2- 2  \beta n_1 \cos{\theta_\mathrm{i}} + 1 }
\end{equation}
and
\begin{equation}
     \cos{\theta_\mathrm{t}} = \frac{ \beta n_1^2 (1- \cos^2{\theta_\mathrm{i}} ) + ( 1- \beta n_1 \cos{\theta_\mathrm{i}} ) \sqrt{  n_2^2( \beta^2 n_1^2  - 2 \beta n_1   \cos{\theta_\mathrm{i}} + 1 )- n_1^2 ( 1-\cos^2{\theta_\mathrm{i}} ) }) } { n_2 (\beta^2 n_1^2- 2  \beta n_1 \cos{\theta_\mathrm{i}} + 1) }\\
\end{equation}
\end{subequations}

\begin{figure}[h!]
	\centering
	\includegraphics[width=0.8\textwidth]{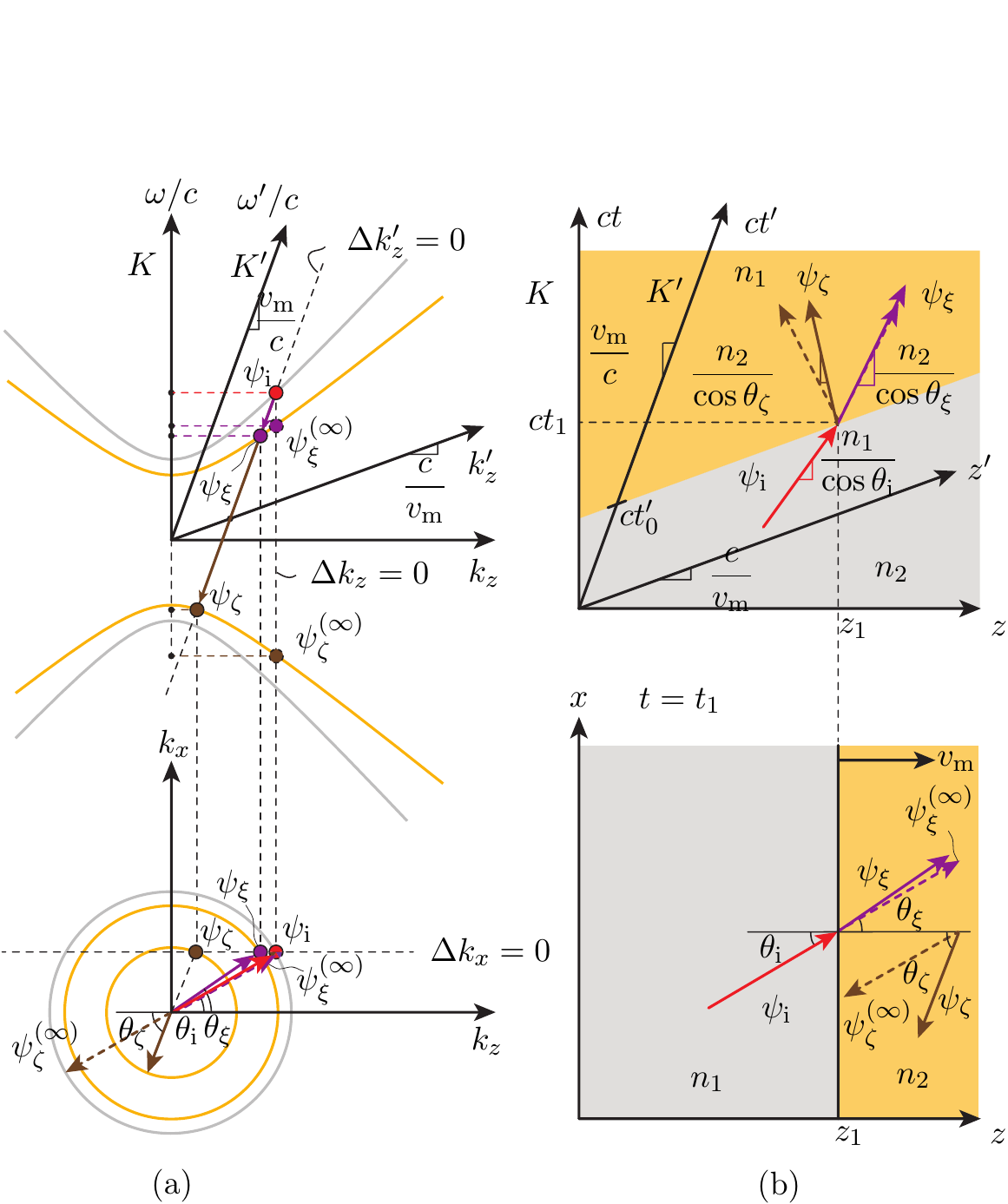}
	\caption{Electrodynamics of USTEM interfaces [Fig.~\ref{fig:GSTEMs_persp}(c)] for the case of oblique incidence [generalization of the normal-incidence case in Fig.~\ref{fig:STEM_interf}(b)] in the superluminal regime. The red, purple and brown arrows correspond to the incident, later-forward and later-backward waves, respectively, and the dashed arrows correspond to the time-interface case where $v=\infty$. (a)~Direct spacetime. (b)~Inverse spacetime.}
	\label{fig:interf_obl_sup}
\end{figure}

Figure~\ref{fig:interf_obl_sup} shows the counterpart of Fig.~\ref{fig:interf_obl_sub} for the case of a superluminal modulation. Similar steps as those leading to Fig.~\ref{fig:interf_obl_sup}(a) are applied, with some variations:
\begin{enumerate}\vspace{-2mm}
    \item plot the hyperbolas of the two media for a given $k_{x\mathrm{i}}$;
    \item plot the axes of the laboratory ($K$) frame, and the axes of a frame with velocity $v=c^2/v_\mathrm{m}$, where $v_\mathrm{m}$ is the modulation velocity. Explanations for this choice are provided in~\cite{Deck_PRB_2018}, but, essentially, choosing this velocity make the primed frame see the interface as a time-interface, i.e., an instantaneous transition;
    \item apply the continuity of the wavevector in the primed frame, i.e. $\Delta k_z'=0$;
    \item project onto the wavevector diagram, setting $\Delta k_x=0$ as previously.
\end{enumerate}\vspace{-2mm}

The results are then translated into the direct space-time diagrams of Fig.~\ref{fig:interf_obl_sup}(b).\\

The mathematical expressions for the deflected angles are found by inserting the Lorentz transformations of the wavenumbers to the  continuity of the wavevectors, $k_{x\mathrm{i}}'=k_{x\mathrm{r}}'=k_{x\mathrm{t}}'$ and $k_{z\mathrm{i}}'=k_{z\mathrm{r}}'=k_{z\mathrm{t}}'$ where the primed frame is moving at a velocity $v=c^2/v_\mathrm{m}$. After a few manipulations, we find
\begin{subequations}\label{sup angular relation}
    \begin{equation}
     \cos{\theta_\xi} = \frac{ \beta n_1^2 (1- \cos^2{\theta_\mathrm{i}} ) + ( \beta - n_1 \cos{\theta_\mathrm{i}} ) \sqrt{  n_2^2( \beta^2  - 2 \beta n_1 \cos{\theta_\mathrm{i}} + n_1^2) - n_1^2 \beta^2 ( 1-\cos^2{\theta_\mathrm{i}} ) }) } { n_2 (\beta^2 - 2  \beta n_1 \cos{\theta_\mathrm{i}} + n_1^2))}
    \end{equation}
    and
    \begin{equation}
   \cos{\theta_\zeta} = \frac{ -\beta n_1^2 (1- \cos^2{\theta_\mathrm{i}} ) + ( \beta - n_1 \cos{\theta_\mathrm{i}} ) \sqrt{  n_2^2( \beta^2  - 2 \beta n_1 \cos{\theta_\mathrm{i}} + n_1^2) - n_1^2 \beta^2 ( 1-\cos^2{\theta_\mathrm{i}} ) }) } { n_2 (\beta^2 - 2  \beta n_1 \cos{\theta_\mathrm{i}} + n_1^2)}.
    \end{equation}
\end{subequations}

Full-wave (FDTD) animations related to this section are provided in the appended file \textbf{GSTEM\_Interface\_scattering.mp4}.


\section{Properties of (UV) Moving Matter}\label{sec:propr_mov_mat}
\subsection{Constitutive Relations}\label{sec:mov_mat_const_rel}

We assume an isotropic and nondispersive medium in uniform motion with the constant rectilinear velocity \mbox{$\ve{v}=v\ve{\hat{z}}$}. In the rest frame of the medium ($K'$), the constitutive relations are
\begin{equation}\label{eq:D'}
\ve{D}'=\epsilon'\ve{E}'\qquad\text{and}\qquad \ve{B}'=\mu' \ve{H}'.
\end{equation}

The corresponding relations in the laboratory frame ($K$) are found by applying the Lorentz transformations~\eqref{eq:field_transf} to the fields in~\eqref{eq:D'} and rearranging the result so as to express $\ve{D}$ and $\ve{B}$ in terms of $\ve{E}$ and $\ve{H}$. This results into the bianisotropic relations~\cite{Kong_EWT_2008}
\begin{subequations}\label{eq:const_rel_bian}
\begin{equation}\label{eq:const_D}
\ve{D}=\te{\epsilon}\cdot \ve{E}+\te{\xi}\cdot\ve{H},
\end{equation}
\begin{equation}\label{eq:const_B}
\ve{B}=\te{\mu} \cdot\ve{H}+\te{\zeta}\cdot\ve{E},
\end{equation}
\end{subequations}
where
\begin{subequations}
\begin{equation}
\te{\epsilon}=
\epsilon'\begin{bmatrix}
\alpha & 0 & 0\\
0 & \alpha & 0\\
0 & 0 & 1
\end{bmatrix}, \quad
\te{\mu}=
\mu'\begin{bmatrix}
\alpha & 0 & 0\\
0 & \alpha & 0\\
0 & 0 & 1
\end{bmatrix},
\quad
\te{\xi}=
\begin{bmatrix}
0 & \chi/c & 0\\
-\chi/c & 0 & 0\\
0& 0 & 0
\end{bmatrix}
\quad\text{and}
\quad \te{\zeta}=-\te{\xi},
\end{equation}
with
\begin{equation}\label{eq:alpha_chi}
\alpha=\frac{1-\beta^2}{1-\beta{n'}^2}
\qquad\text{and}\qquad
\chi=\beta\frac{1-{n'}^2}{1-\beta^2{n'}^2}\qquad(\beta=v/c).
\end{equation}
\end{subequations}

Note that, according to~\eqref{eq:const_rel_bian}, the response fields ($\ve{B}$ and $\ve{D}$) are not parallel to the excitation fields ($\ve{E}$ and $\ve{H}$) in a bianisotropic (and even homoanisotropic, i.e., without magnetoelectric coupling) medium. As a result, the wave propagating in such media are \emph{inhomogenous}, with the phase velocity ($\|\ve{D}\times\ve{B}$) and group velocity  ($\|\ve{S}=\ve{E}\times\ve{H}$) being nonparallel. 

Also note that if the medium at rest is a natural medium (as opposed to a metamaterial) and is not subjected to any external force (e.g., external magnetizing field) or operating in the nonlinear regime~\cite{Caloz_PRAp_10_2018}, then it is necessarily nonmagnetic, so that $\mu'\approx\mu_0$.

\subsection{Inference of First-Order Bianisotropy from the Lorentz Force}\label{sec:force_bian_inf}
The (exact) bianisotropic relations~\eqref{eq:const_rel_bian} were derived mathematically, using Lorentz transformations, with little insight into the related physics. We provide here an original, first-order derivation of these relations, which shows bianisotropy (or R\"{o}ntgen magneto-electric coupling) is an effect that is dominated by the action of the Lorentz force.

\begin{figure}[h!]
	\centering
	\includegraphics[width=0.65\textwidth]{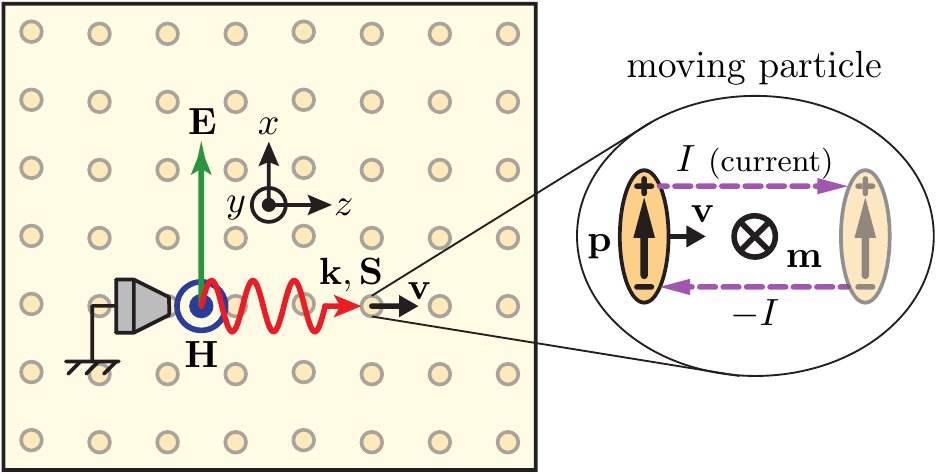}
	\caption{\small ``Bianisotropization'' of matter induced by motion via the Lorentz force.}
	\label{fig:bianisotropization}
\end{figure}

The Lorentz force reads~\cite{Jackson_1998}
\begin{subequations}
	\begin{equation}
	\ve{F}=\ve{F}_E+\ve{F}_H,
	\end{equation}
	where
	\begin{equation}
	\ve{F}_E=q\ve{E}\quad\text{and}\quad\ve{F}_H=q\ve{v}\times\ve{\mu_0H},
	\end{equation}
\end{subequations}
and induces the forces illustrated in Fig.~\ref{fig:bianisotropization} on the atoms and molecules of matter, with the electric part, $\ve{F}_E$, elongating the particle to produce the electric dipole moment response $\ve{p}$, and the magnetic part, $\ve{F}_H$, forming a current loop to produce the manetic dipole moment response $\ve{m}$, which are both represented in the inset of the figure.

The different components of the so-induced electric and magnetic dipole moments, $p_u^{\mathcal{S}_v}$ and $m_u^{\mathcal{S}_v}$ ($u,v=x,y,z$), respectively corresponding to the response $\mathcal{\boldsymbol{R}}=\ve{D},\ve{B}$ to the excitation $\mathcal{\boldsymbol{S}}=\ve{E},\ve{H}$, result at the macroscopic level into susceptibility components $\chi_{uv}^{\mathcal{RS}}$, which can be concisely deduced as follows:
\begin{equation}\label{eq:Lorentz_sys}
\left\{\begin{array}{rll}
E_x\overset{F_E}{\longrightarrow}p_x^{E_x}
& 
\hspace{-0.4cm}
\begin{array} {l}
    \nearrow 
    \vspace{-3pt}
    \\
    \vspace{-3pt}
    \underset{\textcolor{blue}{\beta}}{\searrow}
\end{array} 
&
\hspace{-0.5cm}
\begin{array} {l}
    \chi_{xx}^{DE}\rightarrow \epsilon_{xx}(=\epsilon') \quad \textcolor{blue}{\mathcal{O}(1)}
    \vspace{5pt}
    \\ 
    \vspace{5pt}
    \begin{array} {ccc}
    \hspace{-0.2cm}
    (-)m_y^{E_x}\rightarrow(-)\chi_{yx}^{BE}\rightarrow(-)\zeta_{yx}\propto\beta \quad \textcolor{blue}{\mathcal{O}(\beta)}
    \end{array}       
\end{array}
\vspace{5pt}
\\
\vspace{5pt}
H_y\overset{F_H\propto\textcolor{blue}{\beta}}{\longrightarrow}(-)p_x^{H_y}

& 
\hspace{-0.4cm}
\begin{array} {l}
    \nearrow
    \vspace{-3pt}
    \\
    \vspace{-3pt}
    \underset{\textcolor{blue}{\beta}}{\searrow}
\end{array} 

& 
\hspace{-0.5cm}
\begin{array} {l}
    (-)\chi_{xy}^{DH}\rightarrow(-)\xi_{xy}\propto\beta \quad \textcolor{blue}{\mathcal{O}(\beta)}
    \vspace{5pt}
    \\ 
    \vspace{5pt} 
    (-)(-)m_y^{H_y}\rightarrow\chi_{yy}^{BH}\rightarrow\mu_{yy} \quad \textcolor{blue}{\mathcal{O}(\beta^2)}
\end{array}
\vspace{5pt}
\\
\hline
\\
\vspace{5pt}
E_y\overset{F_E}{\longrightarrow}p_y^{E_y}

& 
\hspace{-0.4cm}
\begin{array} {l}
    \nearrow
    \vspace{-3pt}
    \\
    \vspace{-3pt}
    \underset{\textcolor{blue}{\beta}}{\searrow}
\end{array} 

& 
\hspace{-0.5cm}
\begin{array} {l}
    \chi_{yy}^{DE}\rightarrow \epsilon_{yy}(=\epsilon') \quad \textcolor{blue}{\mathcal{O}(1)}
    \vspace{5pt}
    \\ 
    \vspace{5pt}
    \begin{array} {ccc}
    \hspace{-0.2cm}
    m_x^{E_y}\rightarrow\chi_{xy}^{BE}\rightarrow\zeta_{xy}\propto\beta \quad \textcolor{blue}{\mathcal{O}(\beta)}
    \end{array}       
\end{array}
\vspace{5pt}
\\
\vspace{5pt}
(-)H_x\overset{F_H\propto\textcolor{blue}{\beta}}{\longrightarrow}(-)(-)p_y^{H_x}

& 
\hspace{-0.4cm}
\begin{array} {l}
    \nearrow
    \vspace{-3pt}
    \\
    \vspace{-3pt}
    \underset{\textcolor{blue}{\beta}}{\searrow}
\end{array} 

& 
\hspace{-0.5cm}
\begin{array} {l} 
    \chi_{yx}^{DH}\rightarrow\xi_{yx}\propto\beta \quad \textcolor{blue}{\mathcal{O}(\beta)}
    \vspace{5pt}
    \\ 
    \vspace{5pt} m_x^{H_x}\rightarrow \chi_{xx}^{BH}\rightarrow \mu_{xx} \quad \textcolor{blue}{\mathcal{O}(\beta^2)}
\end{array}

\end{array}\right\},
\end{equation}

where the field pair $(\ve{E},\ve{H})$ was rotated by $\pi/2$ about the $z$ direction in the bottom relations. Ignoring in~\eqref{eq:Lorentz_sys} the terms in $\mathcal{O}(\beta^2)$, whose effect is of second order and whose exact form is also too complex to establish without an \emph{exact} relativistic treatment, we still, remarkably, retrieve the first-order expression of~\eqref{eq:const_rel_bian}, corresponding to the approximation $[\alpha,\chi]=[1,\beta]$. This reveals that bianisotropy is, to the first-order, a direct result of the Lorentz force.


\section{Dispersion Relations of USTEM Structures}\label{sec:USTEM_disp_rel}

\subsection{Dispersion Relation for a Bianisotropic Medium}\label{sec:Mov_Mat_disp_rel}

The dispersion relation for a bianisotropic medium is found by solving the equation~\cite{Kong_EWT_2008}
\begin{equation}\label{eq:general_de}
    \left|\omega^2\te\epsilon-(\te{\ve{k}}+\omega\te\xi)\cdot{\te\mu}^{-1}\cdot(\te{\ve{k}}-\omega\te\zeta)\right|=0.
\end{equation}
%
%

\subsection{Dispersion Relation for a Crystal in the Moving Frame (K')}\label{sec:cryst_disp_rel_Kp}
We shall homogenize a space-time modulated 1+1D crystal with modulation in the $+z$ direction. In the frame that is co-moving with the modulation, $K'$, the interfaces of the crystal are stationary, so that they will be subject to the standard boundary conditions. On the other hand, the media between the interfaces are moving, in the $-z$ direction, and will therefore have the bianisotropic properties given in Supp. Mat~\ref{sec:mov_mat_const_rel}.

In $K'$, the (standard) continuity conditions for an s-polarized wave are
\begin{subequations}
\begin{equation}\label{eq:spol_cont}
    E'_{1y}=E'_{2y}=E_y', \quad H'_{1x}=H'_{2x}=H_x', \quad B'_{1z}=B'_{2z}=B_z',
\end{equation}
and those for a p-polarized are
\begin{equation}\label{eq:ppol_cont}
    H'_{1y}=H'_{2y}=H'_y, \quad E'_{1x}=E'_{2x}=E'_x, \quad D'_{1z}=D'_{2z}=D'_z.
\end{equation}
\end{subequations}

We can find the constitutive relations for the homogenized structure by i)~assuming that the quantities conserved at the interfaces [Eqs.~\eqref{eq:spol_cont} and~\eqref{eq:ppol_cont}] do not depend on the position ($z$), but are rather constant, throughout the structure, as must be the case in the homogeneous regime, ii)~forming  weighted averages, in terms of spacetime travel lengths, $\ell_m'$ ($n=1,2$) and $\ell_\mathrm{t}'=\ell_1'+\ell_2'$ (Fig.~\ref{fig:GSTEM_weighting})~\cite{Deck_APH_10_2019}, of the other quantities , and iii)~using the bianisotropic relations~\eqref{eq:const_rel_bian} for the fields $\ve{D}$ and $\ve{B}.$

For s-polarized waves, we have
\begin{subequations}
\begin{equation}
\begin{split}
    D'_{y\tx{av}}&=\frac{D'_{y1}\ell_1'+D'_{y2}\ell_1'}{\ell_\mathrm{t}'}\\
    &=\frac{\alpha'_1\ell_1'\epsilon_1 E'_{y1}-\chi'_1\ell_1'/c H_{1x}+\alpha_2\ell_2'\epsilon_2 E'_{y2}-\chi'_2\ell_2'/c H'_{2x}}{\ell_\mathrm{t}'}\\
    &=\frac{\alpha'_1\ell_1'\epsilon_1+\alpha'_2\ell_2'\epsilon_2}{\ell_\mathrm{t}'}E'_y
    -\frac{\chi'_1\ell_1'+\chi'_2\ell_2'}{\ell_\mathrm{t}'c}H'_x,
    \end{split}
\end{equation}
\begin{equation}
\begin{split}
    B'_{x\tx{av}}&=\frac{B'_{x1}\ell_1'+B'_{x2}\ell_1'}{\ell_\mathrm{t}'}\\
    &=\frac{\alpha'_1\ell_1'\mu_1 H_{x1}-\chi'_1\ell_1'/c E_{1y}+\alpha'_2\ell_2'\mu_2 H_{x2}-\chi'_2\ell_2'/c E_{2y}}{\ell_\mathrm{t}'}\\
    &=\frac{\alpha'_1\ell_1'\mu_1+\alpha'_2\ell_2'\mu_2}{2}H_x-\frac{\chi'_1\ell_1'+\chi'_2\ell_2'}{2\ell_\mathrm{t}'}E_y,
    \end{split}
\end{equation}
\begin{equation}
\begin{split}
    H'_{z\tx{av}}&=\frac{H'_{z1}\ell_1'+H'_{z2}\ell_2'}{\ell_\mathrm{t}'}\\
    &=\frac{B'_{z}}{\ell'_t}\left(\frac{\ell_1'}{\mu_1}+\frac{\ell_2'}{\mu_2}\right),
    \end{split}
\end{equation}
\end{subequations}
where 
\begin{equation}
 \alpha_m'=\frac{1-v^2/c^2}{1-v^2n_m^2/c^2} \quad\text{and}\quad \chi_m'=-\frac{v}{c}\frac{1-n_m^2}{1-v^2n_m^2/c^2},  
\end{equation}
and for the p-polarized wave
\begin{subequations}
\begin{equation}
\begin{split}
    B'_{y\tx{av}}&=\frac{B'_{y1}\ell_1'+B'_{y2}\ell_2'}{\ell_\mathrm{t}'}\\
    &=\frac{\alpha'_1\ell_1'\mu_1 H'_{y1}+\chi'_1\ell_1'/c E'_{1x}+\alpha'_2\ell_2'\mu_2 H'_{y2}-\chi'_2\ell_2'/c E'_{2x}}{\ell_\mathrm{t}'}\\
    &=\frac{\alpha'_1\ell_1'\mu1+\alpha'_2\ell_1'\mu_2}{\ell_\mathrm{t}'}H'_y
    +\frac{\chi'_1\ell_1'+\chi'_2\ell_2'}{\ell_\mathrm{t}'c}E'_x,
    \end{split}
\end{equation}
\begin{equation}
\begin{split}
    D'_{x\tx{av}}&=\frac{D'_{x1}\ell_1'+D'_{x2}\ell_2'}{\ell_\mathrm{t}'}\\
    &=\frac{\alpha'_1\ell_1'\epsilon_1 E'_{x1}+\chi'_1\ell_1'/c H_{1y}+\alpha'_2\ell_2'\epsilon_2 E_{x2}+\chi'_2\ell_2'/c H'_{2y}}{\ell_\mathrm{t}'}\\
    &=\frac{\alpha'_1\ell_1'\epsilon_1+\alpha'_2\ell_2'\epsilon_2}{\ell_\mathrm{t}'}E_x+\frac{\chi'_1\ell_1'+\chi'_2\ell_2'}{\ell_\mathrm{t}'c}H'_y,
    \end{split}
\end{equation}
\begin{equation}
\begin{split}
    E'_{z\tx{av}}&=\frac{E'_{z1}\ell_1'+E'_{z2}\ell_2'}{\ell_\mathrm{t}'}\\
     &=\frac{D'_{z}}{\ell_\mathrm{t}'}\left(\frac{\ell_1'}{\epsilon_1}+\frac{\ell_2'}{\epsilon_2}\right).
    \end{split}
\end{equation}
\end{subequations}

We can then read out from the previous relations the sought after average constitutive parameters. They are:
\begin{subequations}
\begin{equation}
\epsilon'_{x\text{av}}=\epsilon'_{y\text{av}}=\epsilon'_{\text{av}}= \frac{\alpha'_1\ell_1'\epsilon_1+\alpha'_2\ell_2'\epsilon_2}{\ell_\mathrm{t}'}, \quad 
\epsilon'_{z\text{av}}=\ell_\mathrm{t}'\left(\frac{\ell_1'}{\epsilon_1}+\frac{\ell_2'}{\epsilon_2}\right)^{-1},
\end{equation}
\begin{equation}
    \mu'_{x\text{av}}=\mu'_{z\text{av}}=\mu'_{\text{av}}=\frac{\alpha'_1\ell_1'\mu_1+\alpha'_2\ell_2'\mu_2}{\ell_\mathrm{t}'}, \quad 
\mu'_{z\text{av}}=\ell_\mathrm{t}'\left(\frac{\ell_1'}{\mu_1}+\frac{\ell_2'}{\mu_2}\right)^{-1},
\end{equation}
and 
\begin{equation}
    \chi'_\text{av}=\frac{\chi'_1\ell_1'+\chi'_2\ell_2'}{\ell_\mathrm{t}'},
\end{equation}
with 
\begin{equation}\label{eq:chi}
\te\xi'_\text{av}=
\begin{bmatrix}
0 &\chi'_\text{av}/c & 0\\
-\chi'_\text{av}/c & 0 & 0\\
0& 0 & 0
\end{bmatrix}\qquad\text{and}\qquad\te\zeta'_\text{av}=-\te\xi'_\text{av}.
\end{equation}
\end{subequations}

The dispersion relation in $K'$ is then found by inserting these results into the primed version of~\eqref{eq:general_de} (matter moves in the $K'$ frame in the modulation problem!), which yields two solutions: for the s-polarization,
\begin{subequations}\label{eq:disp_moving_bilayer_kp}
\begin{equation}\label{eq:disp_moving_bilayer_kp_s}
  \frac{1}{\epsilon'_{\tx{av}} \mu'_{z\tx{av}}}({k_x'}^2+{k_y'}^2)+\frac{(k_z'-\chi'_\tx{av} k_0')^2}{\epsilon'_{\tx{av}} \mu'_{\tx{av}}}= {k_0'}^2,
\end{equation}
which are the ellipses plotted in Fig.~\ref{fig:K_Kp_drag}(a) (for the parameters $\epsilon_1=1.1\epsilon_0$, $\mu_1=\mu_0$, $\epsilon_2=3\epsilon_0$, $\mu_2=3\mu_0$ $v=0.25c$), and for the p-polarization,
\begin{equation}\label{eq:disp_moving_bilayer_kp_p}
  \frac{1}{\epsilon'_{z\tx{av}}\mu'_{\tx{av}}}({k_x'}^2+{k_y'}^2)+\frac{(k_z'-\chi'_\tx{av} k_0')^2}{\epsilon'_{\tx{av}} \mu'_{\tx{av}}}= {k_0'}^2.
\end{equation}
\end{subequations}

\subsection{Dispersion Relation for a Crystal in the Laboratory Frame (K)}\label{sec:cryst_disp_rel_K}
Substituting the direct Lorentz relations~\eqref{eq:spec_relative_frames} into~\eqref{eq:disp_moving_bilayer_kp}, and factoring out $k_z$, we find the dispersion relation of the spacetime crystal in the laboratory frame ($K$) as
\begin{subequations}
\begin{equation}\label{eq:disp_moving_bilayer_k_p}
  {k_x}^2+{k_y}^2
  +\mathcal{C}\left[k_z-k_z^+\right]
\left[k_z-k_z^-\right]=0,
\end{equation}
where 
\begin{equation}
  \mathcal{C}=\frac{{n'}^2_{x\tx{av}}}{{n'}^2_{z\tx{av}}}\frac{(1+\beta\chi'_\tx{av})^2-\beta^2{n'}^2_{z\tx{av}}}{1-\beta^2},
\end{equation}
\begin{equation}
    k_z^\pm=k_0\frac{\beta+\chi'_\tx{av}\pm n'_{z\tx{av}}}{1+\beta(\chi'_\tx{av}\pm n'_{z\tx{av}})}
\end{equation}
\end{subequations}
with 
\begin{subequations}
\begin{equation}
  n'_{z\tx{av}}=\sqrt{\epsilon'_{\tx{av}}\mu'_{\tx{av}}}
\end{equation}
and
\begin{equation}
  n'_{x\tx{av}}=\sqrt{\epsilon'_{\tx{av}}\mu'_{z\tx{av}}} \quad \tx{or } \sqrt{\epsilon'_{z\tx{av}}\mu'_{\tx{av}}},
\end{equation}
\end{subequations}
for s- and p-polarization, respectively. 

\subsection{Spacetime Weighted Averaging}\label{sec:ST_WA}

To gain insight into the dependence of the effective refractive index on the angle, we find the effective refractive index for the pure-downstream and pure-upstream cases $\ve{k}=k_z\hat{\ve{z}}$ and $\ve{k}=-k_z\hat{\ve{z}}$ by a simple geometric argument, using Fig.~\ref{fig:GSTEM_weighting}. After some manipulations, given in~\cite{Deck_APH_10_2019}, we find the expression 

\begin{equation}
n_\text{av}^\pm=\frac{n_1\ell_1+n_2\ell_2\mp vn_1n_2\ell_\mathrm{t}/c}{\ell_1+\ell_2\mp v(n_2 \ell_1+n_1 \ell_2)/c}
\end{equation}
which was numerically verified to correspond exactly to~\eqref{eq:disp_moving_bilayer_k_p}, for the case $k_x=0$, when $\epsilon_1/\mu_1=\epsilon_2/\mu_2$.


\section{Derivations for Sec.~\ref{sec:Rindler_transf}}

\subsection{Eqs.~\eqref{eq:a_ap_rel}}\label{sec:der_dudt_vs_dupdtp}

Writing the velocity addition formula~\eqref{eq:v_addition} as
\begin{equation}\label{eq:vaf_ut}
u=\left(u'+v\right)\left(1+\alpha u'\right)^{-1},
\;\text{with}\
\alpha=\beta/c=v/c^2,
\end{equation}
taking the time-derivative of this relation with respect to $t$, using the local relation $dt=\gamma\left(1+\alpha u\right)dt'$ following from Eq.~\eqref{eq:relative_frames}, and setting $u'=0$, yields
\begin{subequations}\label{eq:dudt_loc}
	\begin{align}
		\frac{du}{dt}
		&=\frac{du'}{dt}\left(1+\alpha u'\right)^{-1}+\left(u'+v\right)(-1)\left(1+\alpha u'\right)^{-2}\alpha\frac{du'}{dt} \nonumber \\
		&=\frac{du'}{dt}\left(1+\alpha u'\right)^{-1}\left[1-\left(u'+v\right)\left(1+\alpha u'\right)^{-1}\alpha\right] \nonumber \\
		&=\frac{du'}{\gamma\left(1+\alpha u'\right)dt'}\left(1+\alpha u'\right)^{-1}\left[1-\left(u'+v\right)\left(1+\alpha u'\right)^{-1}\alpha\right] \nonumber \\
		&\overunderset{u'=0}{u=v}{=}\frac{du'}{dt'}\frac{1}{\check{\gamma}}\left(1-v\alpha\right) \nonumber \\
		&=\frac{du'}{dt'}\frac{1}{\check{\gamma}}\left(1-\beta^2\right) \nonumber \\
		&=\frac{du'}{dt'}\frac{1}{\check{\gamma}^3}
	\end{align}
	with
	\begin{equation}
		\check{\gamma}=\frac{1}{\sqrt{1-(u/c)^2}},
	\end{equation}
\end{subequations}
where we have introduced the notation $\check{\gamma}$ to distinguish this $u$-dependent form of $\gamma$ from its original $v$-dependent from in Eqs.~\eqref{eq:gamma} in preparation for later developments.

Note that, by definition, $du/dt=a$ and $du'/dt'=a'=a_0'$ (constant) are the accelerations in $K$ and $K'$, respectively.

\subsection{Eqs.~\eqref{eq:nu_uz_vs_t}}\label{sec:der_nu_uz_vs_t}
Integrating Eq.~\eqref{eq:a_ap_rel}, we find
\begin{subequations}
	\begin{equation}
		\int_v^u\frac{d\tilde{u}}{\left[1-\left(\frac{\tilde{u}}{c}\right)^2\right]^{3/2}}
		=\int_0^ta_0'd\tilde{t},
	\end{equation}
	or
	\begin{equation}
		\frac{u}{\sqrt{1-\left(\frac{u}{c}\right)^2}}-\frac{v}{\sqrt{1-\left(\frac{v}{c}\right)^2}}
		=a_0't,
	\end{equation}
	or yet, 
	\begin{equation}
		\frac{u}{\sqrt{1-\left(\frac{u}{c}\right)^2}}-\frac{v}{\sqrt{1-\beta^2}}
		=a_0't,
	\end{equation}	
	and then, solving for $u$,
	\begin{equation}\label{eq:der_ut}
		u(t)=\frac{a_0't+\gamma v}{\sqrt{1+\left(\frac{a_0't}{c}+\gamma\beta\right)^2}},
	\end{equation}	
\end{subequations}
where the lower integration limits have been chosen so that $u(0)=v$, and $\gamma$ and $\beta$ are given by Eq.~\eqref{eq:gamma}. Substituting then $u(t)=dz/dt$ into the last expression, multiplying the result by $dt$ and further integrating gives
\begin{subequations}
	\begin{equation}
	\int_0^zd\tilde{z}=\int_0^t\frac{a_0'\tilde{t}+\gamma\beta}{\sqrt{1+\left(\frac{a_0'\tilde{t}}{c}+\gamma\beta\right)^2}}d\tilde{t},
	\end{equation}
	or
	\begin{equation}
	z(t)
	=\frac{c^2}{a_0'}\left[\sqrt{1+\left(\frac{a_0't}{c}+\gamma\beta\right)^2}-\sqrt{1+\left(\gamma\beta\right)^2}\right],
	\end{equation}
\end{subequations}
where the lower integration limit has been chosen so that $z(t=0)=0$.

\subsection{Eq.~\eqref{eq:t_vs_tp_KM}}\label{sec:der_loc_KM_transf}
Taking the differential of $t$ Eq.~\eqref{eq:Lorentz_transf_rel} yields
\begin{equation}\label{eq:Lorentz_t_diff}
\begin{split}
dt
&=\gamma\left(dt'+\frac{\beta}{c}dz'\right) \\
&=\gamma\left(1+\frac{\beta}{c}\frac{dz'}{dt'}\right)dt' \\
&=\gamma\left(1+\frac{\beta}{c}u'\right)dt',
\end{split}
\end{equation}
which locally reduces, with $u'=0$, to 
\begin{equation}
dt=\gamma dt',
\end{equation}
and which further reduces then, given that $v=u$, to
\begin{equation}
dt=\check{\gamma}dt',
\;\text{with}\;
\check{\gamma}=\frac{1}{\sqrt{1-\left(\frac{u}{c}\right)^2}}.
\end{equation}
Allowing now $u$ to vary in time, i.e., $u\rightarrow u(t)$, locally extends this relation to 
\begin{equation}
dt=\check{\gamma}(t)dt',
\;\text{with}\;
\check{\gamma}(t)=\frac{1}{\sqrt{1-\left(\frac{u(t)}{c}\right)^2}},
\end{equation}
where the substitution of Eq.~\eqref{eq:nu_u_vs_t} for $u(t)$ leads to
\begin{equation}
\begin{split}
dt
&=\sqrt{1+\left(\frac{a_0't}{c}+\gamma\beta\right)^2}dt',\quad\text{or} \\
\frac{dt}{\sqrt{1+\left(\frac{a_0't}{c}+\gamma\beta\right)^2}}&=dt'.
\end{split}
\end{equation}

Finally, integrating this relation as
\begin{equation}
\int_0^t\frac{d\tilde{t}}{\sqrt{1+\left(\frac{a_0'\tilde{t}}{c}+\gamma\beta\right)^2}}
=\int_0^{\tau}d\tilde{t}'
\end{equation}
gives
\begin{equation}\label{t_vs_tp_KM_app}
t=\frac{c}{a_0'}\sinh\left[\frac{a_0'\tau}{c}+\sinh^{-1}\left(\gamma\beta\right)\right]-\frac{c}{a_0'}\gamma\beta,
\end{equation}
where the lower limits of the integrals have been chosen so that $t(\tau=0)=0$. 

\subsection{Eqs.~\eqref{eq:KM_transf}}\label{sec:der_KM_transf_app}

Substituting Eq.~\eqref{eq:t_vs_tp_KM} into Eq.~\eqref{eq:nu_u_vs_t} yields
\begin{equation}\label{eq:Lorentz_t_diff}
\begin{split}
u(\tau)
&=\frac{c\sinh\left[\frac{a_0'\tau}{c}+\sinh^{-1}\left(\gamma\beta\right)\right]\cancel{-c\gamma\beta+\gamma v}}{\sqrt{1+\left\{\sinh\left[\frac{a_0'\tau}{c}+\sinh^{-1}\left(\gamma\beta\right)\right]\cancel{-\gamma\beta+\gamma\beta}\right\}^2}} \\
&=\frac{c\sinh\left[\frac{a_0'\tau}{c}+\sinh^{-1}\left(\gamma\beta\right)\right]}{\cosh\left[\frac{a_0'\tau}{c}+\sinh^{-1}\left(\gamma\beta\right)\right]} \\	
&=c\tanh\left[\frac{a_0'\tau}{c}+\sinh^{-1}\left(\gamma\beta\right)\right]
\end{split}
\end{equation}
This relation parametrizes Eqs.~\eqref{eq:gamma} as
\begin{subequations}\label{eq:beta_gamma_par}
	\begin{equation}
	\check{\beta}(\tau)=\frac{u(\tau)}{c}=\tanh\left[\frac{a_0'}{c}\tau+\sinh^{-1}\left(\gamma\beta\right)\right]
	\end{equation}
	and
	\begin{equation}
	\check{\gamma}(\tau)=\frac{1}{\sqrt{1-\beta^2(\tau)}}=\cosh\left[\frac{a_0'}{c}\tau+\sinh^{-1}\left(\gamma\beta\right)\right],
	\end{equation}	
\end{subequations}
where the accents $\check{}$ in the left-hand sides have been used to distinguish the $\tau$-parametric functions from their $\tau=0$ counterpart in the right-hand side, given by Eqs.~\eqref{eq:gamma}.

Inserting now these relations into Eqs.~\eqref{eq:nu_z_vs_t} and~\eqref{eq:t_vs_tp_KM} provides the spacetime coordinates of the local Lorentzian coordinate systems along $c\tau$ as
\begin{subequations}\label{eq:zct_Op_tau}
	\begin{equation}
		\begin{split}
			z_{O'}(\tau)
			&=\frac{c^2}{a_0'}\left\{\sqrt{1+\sinh^2\left[\frac{a_0'}{c}\tau+\sinh^{-1}\left(\gamma\beta\right)\right]}-\gamma\right\} \\
			&=\frac{c^2}{a_0'}\left\{\cosh\left[\frac{a_0'}{c}\tau+\sinh^{-1}\left(\gamma\beta\right)\right]-\gamma\right\}
		\end{split}
	\end{equation}
	\begin{equation}
		ct_{O'}(\tau)=\frac{c^2}{a_0'}\left\{\sinh\left[\frac{a_0'\tau}{c}+\sinh^{-1}\left(\gamma\beta\right)\right]-\gamma\beta\right\}.
	\end{equation}
\end{subequations}

Finally, the sought after transformation relation are found as by substituting Eqs.~\eqref{eq:beta_gamma_par} and~\eqref{eq:zct_Op_tau} into the parametrized Lorentz inverse relations with $t'=0$ as
\begin{subequations}
	\begin{equation}
		\begin{split}
			z
			&=z_{O'}(\tau)+\gamma(\tau)z' \\
			&=\left(z'+\frac{c^2}{a_0'}\right)\cosh\left[\frac{a_0'}{c}\tau+\sinh^{-1}\left(\gamma\beta\right)\right]-\frac{c^2}{a_0'}\gamma
		\end{split}
	\end{equation}
	and
	\begin{equation}
		\begin{split}
			ct
			&=ct_{O'}(\tau)+\gamma(\tau)\beta(\tau)z' \\
			&=\left(z'+\frac{c^2}{a_0'}\right)\sinh\left[\frac{a_0'}{c}\tau+\sinh^{-1}\left(\gamma\beta\right)\right]-\frac{c^2}{a_0'}\gamma\beta,
		\end{split}
	\end{equation}
\end{subequations}
where $\tau$ may be ultimately exchanged with $t'$ to give
\begin{subequations}
	\begin{equation}
	\begin{split}
	z
	&=\left(z'+\frac{c^2}{a_0'}\right)\cosh\left[\frac{a_0'}{c}t'+\sinh^{-1}\left(\gamma\beta\right)\right]-\frac{c^2}{a_0'}\gamma
	\end{split}
	\end{equation}
	and
	\begin{equation}
	\begin{split}
	ct
	&=\left(z'+\frac{c^2}{a_0'}\right)\sinh\left[\frac{a_0'}{c}t'+\sinh^{-1}\left(\gamma\beta\right)\right]-\frac{c^2}{a_0'}\gamma\beta.
	\end{split}
	\end{equation}
\end{subequations}

\subsection{Rindler-Kottler-M{\o}ller Metric}\label{sec:KM_metric_sm}

To determine the Rindler-Kottler-M{\o}ller metric, we substitute the Rindler-Kottler-M{\o}ller transformation relations~\eqref{eq:KM_transf} into the Minkowski metric, which gives
\begin{align}
ds^2
&=-\left(cdt\right)^2+dz^2 \nonumber \\
&=-\left[\frac{\partial(ct)}{\partial z'}dz'+\frac{\partial(ct)}{\partial(ct')}d(ct')\right]^2+\left[\frac{\partial z}{\partial z'}dz'+\frac{\partial z}{\partial(ct')}d(ct')\right]^2 \nonumber \\
&=-\left[\sinh(\zeta t'+\xi)dz'+(z'+z_0)\zeta\cosh(\zeta t'+\xi)d(ct')\right]^2 \nonumber \\
&\qquad\qquad+\left[\cosh(\zeta t'+\xi)dz'+(z'+z_0)\zeta\sinh(\zeta t'+\xi))(dct')\right]^2 \nonumber \\
&=-(z'+z_0)^2\zeta^2\left(cdt'\right)^2+dz'^2.
\end{align}
So, the Kottler-M{\o}ller metric is
\begin{subequations}
	\begin{equation}
		ds'^2=-[(z'+z_0)\zeta]^2\left(cdt'\right)^2+dz'^2,
	\end{equation}
	or, generalizing to include the (motion-less) directions $x$ and~$y$,
	\begin{equation}
		ds'^2=-[(z'+z_0)\zeta]^2\left(cdt'\right)^2+r'^2.
	\end{equation}
\end{subequations}

\end{document}